\documentclass[a4paper,fleqn,usenatbib]{mnras}

\usepackage{newtxtext,newtxmath}

\usepackage[T1]{fontenc}
\usepackage{ae,aecompl}

\usepackage{graphicx}	
\usepackage{amsmath}	
\usepackage{amssymb}	

\usepackage{mathtools}
\usepackage{color}

\newcommand{\grb}{GRB\,170817A}
\newcommand{\host}{NGC\,4993}
\newcommand{\fracb}[2]{\left(\frac{#1}{#2}\right)}

\title[Off-Axis Emission from Short GRBs]{Off-Axis Emission of Short GRB Jets from Double Neutron Star Mergers and GRB$\,$170817A}

\author[J. Granot et al.]{Jonathan Granot,$^{1}$\thanks{E-mail: granot@openu.ac.il (JG)},
Ramandeep Gill $^{1,2}$, Dafne Guetta$^{3}$ and Fabio De Colle$^{4}$
\\
$^{1}$Department of Natural Sciences, The Open University of Israel, 1 University Road, POB 808, Raanana 4353701, Israel\\
$^{2}$Physics Department, Ben-Gurion University, P.O.B. 653, Beer-Sheva 84105, Israel\\
$^{3}$ORT-BRAUDE College, Snunit 51, Karmiel, Israel\\
$^{4}$Instituto de Ciencias Nucleares, Universidad Nacional Aut{\'o}noma de M{\'e}xico, A. P. 70-543 04510 D. F. Mexico
}

\date{Accepted XXX. Received YYY; in original form ZZZ}

\pubyear{2017}

\begin{document}
\label{firstpage}
\pagerange{\pageref{firstpage}--\pageref{lastpage}}
\maketitle



\begin{abstract}
The short-duration ($\lesssim2\;$s) \grb\ in the nearby ($D=40\;$Mpc) elliptical galaxy \host\ is 
the first electromagnetic counterpart of the first gravitational wave (GW) detection of a binary 
neutron-star (NS-NS) merger. It was followed by optical, IR, and UV emission from half a day up to 
weeks after the event, as well as late time X-ray and radio emission. The early UV, optical, and IR 
emission showed a quasi-thermal spectrum suggestive of radioactive-decay powered kilonova-like 
emission. Comparison to kilonova models favors the formation of a short-lived ($\sim1\;$s) 
hypermassive NS, which is also supported by 
the $\Delta t\approx1.74\;$s delay between the GW chirp signal and the prompt GRB onset. However, the late onset of the X-ray (8.9$\;$days) 
and radio (16.4$\;$days) emission, together with the low isotropic equivalent $\gamma$-ray energy output 
($E_{\rm\gamma,iso}\approx5\times10^{46}\;$erg), strongly suggest emission from a narrow relativistic jet viewed off-axis. 
Here we set up a general framework for off-axis GRB jet afterglow emission, comparing analytic and numerical approaches, 
and showing their general predictions for short-hard GRBs that accompany binary NS mergers. The prompt GRB emission suggests 
a viewing angle well outside the jet's core, and we compare the afterglow lightcurves expected in such a case to the X-ray 
to radio emission from \grb. We fit an afterglow off-axis jet model to the X-ray and radio data and find that the observations 
are explained by a viewing angle $\theta_{\rm obs}\approx16^\circ-26^\circ$, GRB jet energy $E\sim10^{48.5}-10^{49.5}~{\rm erg}$, 
and external density $n\sim10^{-5}-10^{-1}~{\rm cm}^{-3}$ for a $\xi_e\sim 0.1$ non-thermal electron acceleration efficiency.
\end{abstract}

\section{Introduction}

Gravitational waves were first detected only two years ago, and a century after they were predicted by Albert Einstein, 
by the Advanced Laser Interferometer Gravitational-wave Observatory \citep[LIGO;][]{Abbott+16a}. This first detection 
was from two coalescing black holes (BHs), and it has just led to the award of a Nobel prize (the 2017 prize in physics 
to Rainer Weiss, Barry C. Barish, and Kip S. Thorne). This groundbreaking discovery marked the dawn of a new era of GW 
astronomy. It was quickly followed by two other BH-BH mergers, and recently yet another one that was also detected by 
VIRGO \citep{Abbott+17a}. 
This has clearly established the ability and sensitivity of current GW detectors to robustly and securely detect such sources 
(merging BHs of a few tens of solar masses) out to $\sim$Gpc distances. Moreover,  
LIGO is also capable of detecting GWs from compact binary mergers involving neutron stars (NSs), namely NS-NS and NS-BH, 
at a volume-weighted mean distance of $\sim\,70\;$Mpc and $\sim\,110\;$Mpc, respectively, and has set an upper limit of 
$12,\!600\;{\rm Gpc}^{-3}\,{\rm yr}^{-1}$ on the NS-NS merger rate \citep[90\% CL;][]{Abbott+16b}.

However, no significant electromagnetic emission is generally expected for a BH-BH merger (in most scenarios). 
Nonetheless, its detection is of great importance in NS-NS or NS-BH mergers, which have been suggested to be 
the progenitors of short-hard gamma-ray bursts \citep[SGRBs; e.g.][]{ELPS89,NPP92}. A NS-NS merger is expected 
to lead to the formation of a BH, which is possibly preceded by a short-lived hypermassive neutron star (HMNS). 
Accretion onto the BH then launches a relativistic jet \citep[e.g.][]{Rezzolla11} that eventually reaches bulk 
Lorentz factors $\Gamma\gtrsim100$ and power a SGRB -- a short ($\lesssim2\;$s) intense burst of $\gamma$-rays 
with a typical (peak $\nu F_\nu$) photon energy $E_{\rm pk}\sim400\;$keV and isotropic-equivalent $\gamma$-ray 
energy output $E_{\gamma,\rm iso}\simeq10^{49}-10^{51}\;$erg \citep[][]{Nakar07,Berger14}. GRBs of the long-soft 
class, on the other hand, are known to originate from the death of massive stars, and are found to be   
associated with star-forming regions and type Ic core-collapse supernovae \citep[e.g.,][]{WB06}.


The detection of an SGRB, \grb ~\citep{GCN21520}, in coincidence with the first GW detection of a NS-NS merger 
\citep[][]{Abbott+17b}, in the relatively nearby ($D=40\;$Mpc) elliptical galaxy \host, provides the long awaited 
``smoking gun'' that binary NS mergers indeed give rise to SGRBs. This exciting news has led to a huge multi-wavelength
follow-up effort, involving observations by detectors across the EM spectrum \citep[see for e.g.][and references therein]{capstone}. 
This source was detected 
in the optical, UV, and IR after about half a day, as well as in X-rays (after 8.9$\;$days) and radio (after 16.4$\;$days) 
that took advantage of the much better position measurement from the preceding optical detections.  All these measurements 
provide critical information regarding the jet geometry, merger ejecta, and r-process elements \citep[e.g.][]{RPN13}.

A binary NS merger can produce EM radiation over a wide range of wavelengths and time scales. During a NS-NS 
(or NS-BH) merger some sub-relativistic ejecta (of mass $\sim10^{-3}-10^{-2}\;M_\odot$) is thrown out along 
the orbital plane at a modest fraction of the speed of light, $\beta = v/c \sim 0.1-0.3$. Rapid neutron capture 
in the sub-relativistic ejecta \citep[e.g.,][]{LS76} is hypothesized to produce a kilonova (also known as a 
macronova or mini-supernova) -- an optical and near-infrared signal lasting hours to weeks \citep[e.g.,][]{LP98} 
powered by radioactive decay. Eventually, this sub-relativistic ejecta is decelerated significantly by sweeping up mass 
comparable to its own, and transfers most of its kinetic energy ($\sim10^{50}-10^{51}\;$erg) to the 
swept up shocked ambient medium. The latter produces radio emission through synchrotron radiation, which typically peaks 
on a timescale of months to years after the merger \citep[e.g.,][]{NP11}. If a short-lived ($\lesssim1\;$s) 
HMNS is formed during the merger, then its slightly delayed (relative to the NS-NS merger) collapse triggers 
the formation of a relativistic jet. In this case, the jet needs to bore its way through the neutrino-driven wind 
that was launched before this collapse, thus creating a cocoon whose cooling and radioactive decay signatures 
may produce observable emission in the optical-UV on timescales of an hour or so \citep{GNP17}. The delayed collapse and the time it takes for the jet to bore through the wind may account \citep[e.g.][]{short} for the delay of $1.74\pm0.05\;$s between the GW chirp signal and the GRB prompt emission onset \citep{Abbott2017}.

During a NS-NS merger, the collision between the two neutron stars drives a strong shock into them, which accelerates 
as it reaches the sharp density gradient in their outer layer, thus producing a quasi-spherical ultra-relativistic 
outflow, which may give rise to detectable emission in X-rays (peaking promptly), optical (peaking within seconds) and radio 
(peaking within several hours to a day) \citep{Kyutoku14}. It has also been suggested that several seconds prior to or 
tens of minutes after the merger, one might detect a coherent radio burst lasting milliseconds \citep[e.g.,][]{HL01,Z14}.  

If the relativistic jet that forms after the NS-NS merger happens to point towards us, then we may observe a short 
prompt gamma-ray emission episode from a SGRB, lasting $\lesssim2\;$s, followed by afterglow emission in X-ray, 
optical, and radio lasting for hours, days or weeks \citep[e.g.,][]{ELPS89,NPP92,Nakar07,Berger14,Fong15}. However, 
in most cases the relativistic jet will not point towards us, and in this work we will focus on the observable 
signatures that are expected in this case. 

The search for orphan afterglows: afterglows which are not associated with observed prompt GRB emission, has 
immediately followed the realization that GRBs are beamed with rather narrow opening angles, while the afterglow 
itself could be observed over a wider angular range. \cite{Rhoads97} was the first to suggest that observations of 
orphan afterglows would enable us to estimate the opening angles and the true rate of GRBs. Because of relativistic 
beaming, only a region of angular size $1/\Gamma$ around the line of sight is visible. Therefore, when a jet of 
initial half-opening angle $\theta_0 > 1/\Gamma$ points towards us then initially we cannot notice that it is indeed 
a jet and not part of a spherical flow. This is typically valid during the prompt gamma-ray emission, and  we can 
learn that it is indeed a jet only later during the afterglow stage when the jet decelerates whereby its Lorentz 
factor $\Gamma$ drops below $1/\theta_0$, resulting in an achromatic steepening of the flux decay rate, known as a 
jet break \citep{Rhoads99,SPH99}. 

When a jet is viewed off-axis, from outside of its initial aperture, at a viewing angle $\theta_{\rm obs}>\theta_0$ , then its prompt gamma-ray emission is significantly suppressed and would in most cases be missed, unless the GRB is particulary bright, nearby, or viewed from very close to its edge, 
such that $\Gamma(\theta_{\rm obs}-\theta_0)\lesssim\;$a few. However, the afterglow emission becomes visible once the beaming cone of the afterglow radiation reaches our line of sight. The jet's beaming cone subtends an angle of $\theta_j+1/\Gamma$ from its symmetry axis, the first term ($\theta_j$) is geometrical (since due to relativistic beaming the brightest part of the jet is the one closest to our line of sight) and can grow with time by up to $\theta_j\leq\theta_0+1/\Gamma$ to the extent that the jet expands sideways.  The second term ($1/\Gamma$) arises from relativistic beaming. Altogether, for $\theta_{\rm obs}\gtrsim2\theta_0$ the beaming cone reaches the line of sight when $\Gamma$ drops to $\sim1/\theta_{\rm obs}$, at which the lightcurve peaks, and subsequently approaches that for an on-axis observer.

For a jet sideways expansion at close to its sound speed $\theta_j$ grows exponentially with radius after the jet break radius \citep[][;while noticeable deviations from this scaling are expected for $\theta_j\gtrsim0.1$ \citep{GP12}, the resulting analytic lightcurve scalings are still very useful also for $\theta_{\rm obs}\gtrsim0.1$.]{Rhoads99,SPH99}. 
This results in a post jet break dynamics and therefore lightcurves that are universal, in the sense that they are independent of $\theta_0$, and depend only on the jet's true energy $E$ rather than on its isotropic equivalent energy \citep{off-axis2002}. Here we follow and generalize the result of \citet{NPG02} for the calculation of these post jet break lightcurves and peak time and flux for off-axis observers.

In \S\,\ref{sec:optical} we discuss the possible origins of the IR, optical, and UV emission (\S\,\ref{sec:opt_origin}), focusing on kilonova emission (\S\,\ref{sec:kilonova}) and how it compares to the afterglow emission (\S\,\ref{sec:KNvsAG}), and briefly mention the expected late time radio emission from sub-relativistic ejecta (\S\,\ref{sec:late_radio}). 
The early IR, optical, and UV data emission appear to be dominated by kilonova-like emission, while the late time X-ray and radio emission are most likely off-axis afterglow emission. In \S~\ref{sec:host}, we use the projected distance of the SGRB to show that the binary could not have moved in a straight 
line from its birth site to its merger site. In \S\,\ref{sec:model} we outline a simple analytic model for off-axis afterglow emission, which provides the peak time and flux for lightcurves seen by observers outside of the jet's initial aperture. The general predictions of this model (for $\theta_{\rm obs}\gtrsim2\theta_0$) are presented in \S\,\ref{sec:genres}.
In \S\,\ref{sec:sims} we present numerical lightcurves from relativistic hydrodynamic simulations of a GRB jet during 
the afterglow phase, and compare them to the predictions of the simple analytic model. Next, we directly compare the 
numerical lightcurves from hydrodynamic simulations to the data for \grb~in \S\,\ref{sec:compdata}. Both upper limits 
and flux measurements are taken into account, while ignoring the early IR, optical, and UV data since they are likely 
dominated by kilonova emission. For the purpose of detailed comparison with the data, given the moderately large number 
of model parameters and corresponding parameter space to explore, we use: (i) scaling relations of the dynamical equations 
in order to avoid the need for performing a large number of hydrodynamic simulations, as well as (ii) additional scalings 
of the flux density at the different power-law segments of the afterglow synchrotron spectrum, in order to greatly reduce 
the number of required numerical lightcurve calculations. Our conclusions are discussed in \S\,\ref{sec:diss}.

\section{Interpretation of the Optical and IR Data for \grb}
\label{sec:optical}

\subsection{General Description of the Optical to IR Data and Some Possible Interpretations}
\label{sec:opt_origin}

In this section we consider the optical and IR data reported of the follow up of GW170817 \citep{capstone,Arcavi2017,Chornock2017,Coulter+17} 
and examine possible interpretations in the context of different scenarios for transient electromagnetic (EM) emission that follows 
the NS-NS merger that produced the gravitational wave signal. Several possible types of EM transients associated with NS-NS (or NS-BH) 
mergers have been suggested in the literature, including short-hard GRBs and afterglows viewed off-axis (in radio, optical or X-ray), 
which we have already mentioned, or a kilonova (in IR, optical, UV) that we address in the following subsection.

Another possible source of emission following the NS-NS merger is from the cocoon that is formed as the GRB relativistic jet bores 
its way out of the dynamic outflow from the merger and/or the neutrino-driven wind from a short lived HMNS, as described in 
\citet{GNP17} and \citet{Lazzati+17b}. The emission from the cocoon consists of two components: the cooling emission of the hot cocoon that gives a brief 
($\sim$ 1 hr) blue signal and the cocoon macronova  that arises from radioactive decay within the expanding cocoon material. As can 
be seen in Fig.~3 of \citet{GNP17}, the lightcurves in this model strongly depend on the viewing angle and they may partially fit 
the data. However, the peak generally occurs too early $\sim10^3-10^4\;$s compared to \grb\ observations (where it is at $\sim10^5\;$s), 
and in the UV rather than optical.

It has also been suggested \citep{Kyutoku14} that a rapid and quasi-isotropic emission may occur from the shock that is produced 
as the two NSs collide during their merger. This shock quickly goes through the compact stars' interior and accelerates as it reaches the sharp 
density gradient in their outer layers, producing a quasi-spherical ultra-relativistic outflow. The interaction of this outflow 
with the extrenal medium can produce potentially detectable synchrotron emission from radio to X-ray, peaking within milliseconds 
in X-ray, within seconds in optical, and after a day or so in radio. Since it arises from external shock synchrotron emission it 
has an afterglow-like spectrum, which is very different from the quasi-thermal kilonova spectrum. This might cause some confusion 
with afterglow emission. However it shows quite different temporal properties and carries less energy in ultra-relativistic material 
compared to the GRB jet. Moreover, it should have only a modest dependence on the viewing angle $\theta_{\rm obs}$, while the 
afterglow emission strongly depends on $\theta_{\rm obs}$. 

The optical to IR data of \grb\ from half a day to a week after the event show a quasi-thermal spectrum, which supports a kilonova origin rather and favours it over an afterglow or merger-shock origin. While cocoon emission could also be quasi-thermal, the peak time and temperature are more consistent with a kilonova origin, which will therefore be considered next.


\subsection{Modeling the Kilonova: Inferring the Mass and Velocity of Sub-Relativistic Ejecta}
\label{sec:kilonova}

A kilonova or macronova is a general term for quasi-thermal emission powered by the radioactive decay in sub-relativistic 
neutron-rich material that is ejected during a NS-NS or NS-BH merger. During and following a double NS merger, a small 
fraction of the system's mass ($M_{\rm ej}\sim10^{-4}-10^{-2}M_\odot$ for NS-NS or up to  $10^{-1}M_\odot$ for NS-BH) is 
expected to be ejected at mildly relativistic speeds ($\beta_{\rm ej}\sim0.1-0.3$) and to produce EM emission due to the 
heating by radioactive decay of r-process elements. Models describing this process are still rather uncertain. Below we 
discuss some of the basic properties of such models, and their ability to account for the optical to IR emission in \grb.

Tidal and hydrodynamic interactions are expected to produce a ``dynamical'' ejecta that is highly neutron rich ($Y_e<0.1$). 
The generation of $A>130$ elements (Lanthanides) by r-process nucleosynthesis within this ejecta leads to high opacity 
($\kappa>1\;{\rm cm^2\;g^{-1}}$) and hence to a relatively low peak luminosity, $L_{\rm peak}\sim10^{40}-10^{41}\;{\rm erg\;s^{-1}}$, 
with a temperature in the IR range, $kT_{\rm eff}(t_{\rm peak}) \sim 0.2\;$eV, and a peak time $t_{\rm peak}\sim\;$a few 
days (the so called ``red kilonova''). The spectrum of the kilonova can be fitted with a blackbody spectrum. Fitting this 
spectrum to the data \citep[e.g.][]{Smartt2017} provides how the bolometric luminosity, the black body temperature, and 
radius evolve with time.

The peak time $t_{\rm peak}$ occurs when the photon diffusion time through the ejecta equals the dynamical time, and it 
thus depends on $M_{\rm ej}$, $\beta_{\rm ej}$, and $\kappa$. The peak luminosity, $L_{\rm peak}$, is set by the radioactive 
decay rate at $t_{\rm peak}$, which depends on the ejecta's mass ($M_{\rm ej}$) and composition (Eqs.~(\ref{Lkilopeak}), (\ref{tkilopeak})).



\citet{Barkas2013} have calculated kilonova lightcurves using the time-dependent multi-wavelength radiation transport 
code \verb Sedona. They obtain a wide range of possible peak luminosities, in the range of 
$(0.1\,\!-\,\!4)\times 10^{41}\;{\rm erg\;s^{-1}}$, peaking in the optical and/or IR, depending on model parameters. 

Another example of a kilonova model is that by \citet{KFM}. In this model, a NS-NS merger leads to the formation of a short-lived HMNS, whose lifetime (until it collapses to a black hole), $t_{\rm HMNS}$, affects the properties of the outflow and its observable signatures. They find that the peak emission frequency goes from UV-optical (first $\sim2\;$days) to infrared ($\sim10\;$days). In this scenario, $M_{\rm ej}$ may be enhanced by neutrino-driven wind from the HMNS, and/or after it collapse, from the accretion torus around the newly formed BH. The larger $t_{\rm HMNS}$, the longer the expected neutrino irradiation, and the electron fraction $Y_e$ in the outflow is enhanced. The value of $Y_e$ determines the final composition of the ejecta and has a strong effect on the lightcurve of the kilonova.

\citet{KFM} have performed hydrodynamical wind simulations starting with an equilibrium torus of mass  $0.03M_\odot$ surrounding a central remnant. Different simulations were performed for $t_{\rm HMNS}={0,\,30,\,100,\,300}\;$ms, where for one model the neutron star was assumed to survive indefinitely ($t_{\infty}$).
All models assume a non rotating BH.

From the comparison of the data with the lightcurves of different kilonova models mentioned above, none of these 
models can adequately fit the Optical and IR data \citep{Cowperth} detected from the follow up of GW170817. 
In that work they show that a single component kilonova model does not provide a good fit to the data, and instead 
they use a two component kilonova model. One ``blue'' component that fits the early optical emission is lanthanide 
poor and has $M_{\rm ej}\approx 0.01 M_{\odot}$ and $\beta_{\rm ej}\approx 0.3$, while the second, ``red'' component that 
fits the later optical to IR data is lanthanide-rich with  $M_{\rm ej}\approx 0.04 M_{\odot}$ and $\beta_{\rm ej}\approx 0.1$.
These results are consistent with simple analytic estimates for the peak luminosity and time \citep{MF14},
\begin{eqnarray}\label{Lkilopeak}
L_{\rm peak} &=&  4.3\times 10^{41} \beta_{-1}^{1/2} M_{-2}^{1/2}\kappa_{0}^{-1/2}\;{\rm erg\; s^{-1}}\ , 
\\ \label{tkilopeak}
t_{\rm peak} &=&  1.4\beta_{-1}^{-1/2} M_{-2}^{1/2}\kappa_{0}^{1/2}\;{\rm days}\ ,
\end{eqnarray}
where $\beta_{-1}=\beta_{\rm ej}/0.1$, $M_{-2}=M_{\rm ej}/10^{-2}M_\odot$, and the ejecta's opacity is $\kappa = \kappa_0\;{\rm cm^{2}\;g^{-1}}$.

\subsection{Comparison of the Optical and IR Data to Kilonova Vs. Afterglow Lightcurves}
\label{sec:KNvsAG}

The observed spectrum in the optical to IR range appear to be quasi-thermal. This favors a kilonova-like emission, powered by radioactive decay in a Newtonian neutron rich ejecta from the NS-NS merger. Such a spectrum is not expected for the afterglow emission, which consists of smoothly joining power-law segments \citep{SPN98,GS02}. 

\begin{figure}
\centering
\includegraphics[width=0.99\columnwidth]{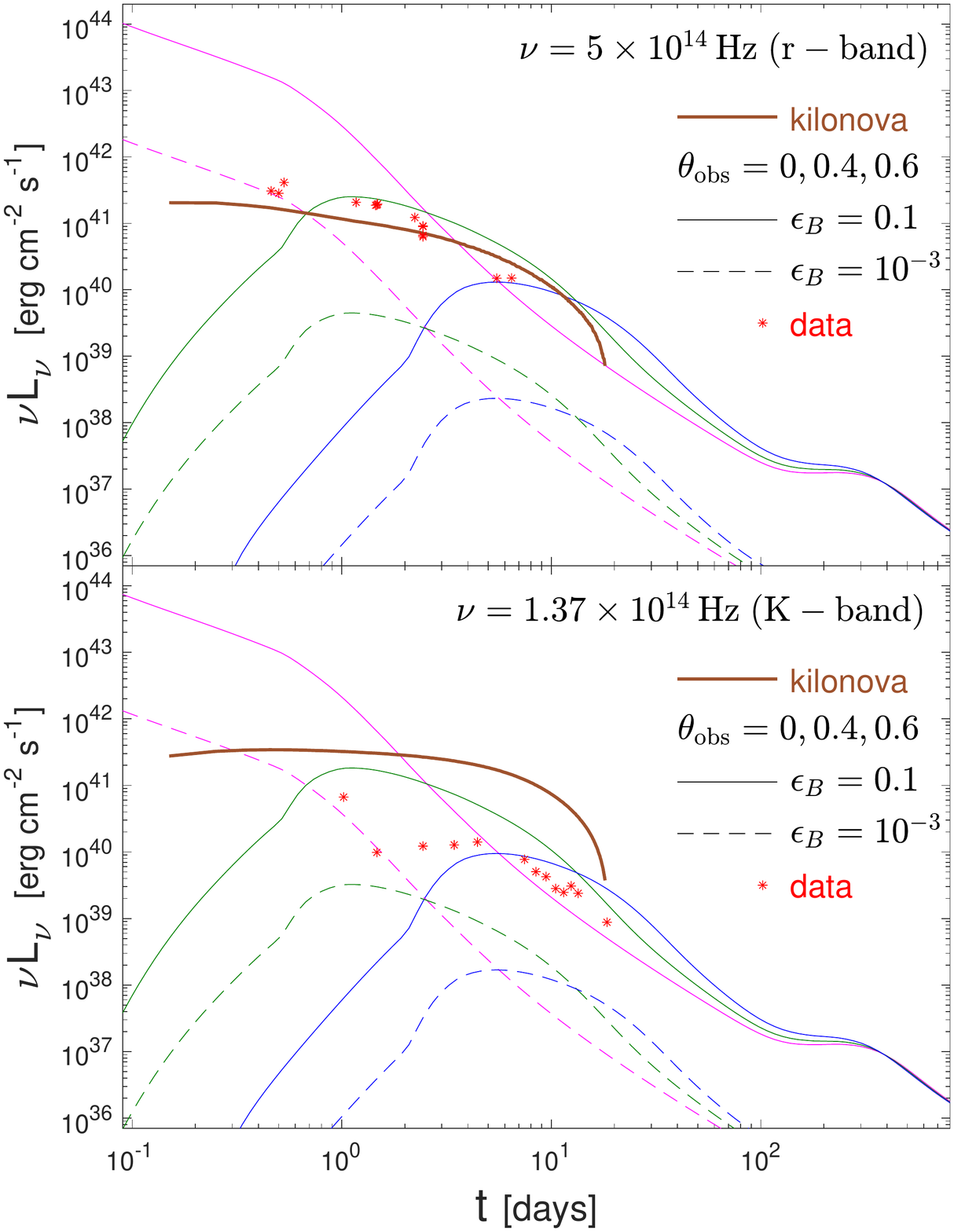} \caption{
Comparison of optical ({\it upper panel}) and IR ({\it lower panel}) data ({\it red asterisks}) for \grb\ to theoretical models \citep[][with $t_{\rm HMNS}=300\;$ms]{KFM} for emission from a kilonova ({\it brown thick line}) and afterglow emission from a GRB jet observed from different viewing angles ($\theta_{\rm obs}=0,\,0.4,\,0.6$ in magenta, dark green, and blue, respectively) and for two different values of the magnetic field equiprtition value ($\epsilon_B = 0.1,\,10^{-3}$ in solid and dashed lines, respectively) while the other parameters are fixed to their fiducial values ($E=10^{49}\;$erg, $p=2.5$, $n_0=1$, $\epsilon_e=0.1$, $\theta_0=0.2$), based on relativistic hydrodynamic simulations (see \S\S\,\ref{sec:sims} and \ref{sec:compdata} for details).}
\label{fig:LC_kilonova_off-axis}
\end{figure}

In Fig.~\ref{fig:LC_kilonova_off-axis} we test the consistency of such an interpretation with the data by comparing the data for \grb\ in optical ($\nu=5\times10^{14}\;$Hz or approximately $r$-band) and IR ($\nu=1.37\times10^{14}\;$Hz; $K$-band) to the expected lightcurves from a kilonova and from a GRB jet for different viewing angles $\theta_{\rm obs}$. The data appear to be broadly consistent with a kilonova origin, although a single component model does not provide a good fit to all the data as discussed in the previous subsection. For a wide range of model parameter values the emission from an off-axis GRB jet can easily be dimmer than the kilonova emission near its peak, and such parameters are also required by the upper limits and detections in the radio and X-ray, as discussed is \S\,\ref{sec:compdata}. However, it can become dominant at later times, which could be tested by a careful analysis of the late time lightcurves and spectra. Moreover, it may be more promising to search for the (off-axis) GRB afterglow emission in the radio or X-ray where it could potentially dominate over the kilonova emission even on a timescale of hours to days.

\subsection{The Late Time Radio Emission from the Sub-Relativistic Ejecta}
\label{sec:late_radio}

The emission from the interaction of the sub-relativistic ejecta with the external medium was 
studied by \citet{NP11}. They find that for sufficiently high radio frequencies 
(typically $\nu\gtrsim1\;$GHz) this emission peaks at the deceleration time $t_{\rm dec}$ and 
flux density given by
\begin{eqnarray}
t_{\rm dec} = 0.77E_{49}^{1/3}n_0^{-1/3}(3\beta_0)^{-5/3}\;{\rm yr}\ ,\quad\quad\quad\quad\quad\quad\quad\quad\quad\ \,
\\
F_\nu^{\rm peak} \approx 0.25E_{49}n_0^\frac{p+1}{4}\epsilon_{B,-1}^\frac{p+1}{4}\epsilon_{e,-1}^{p-1}(3\beta_0)^\frac{5p-7}{2}D_{\rm40Mpc}^{-2}\nu_{9.93}^\frac{1-p}{2}\;{\rm mJy}\ ,
\end{eqnarray}
where we have used a fiducial distance of $D=40D_{\rm 40Mpc}\;$Mpc, initial dimensionless 
velocity $\beta_0=1/3$, and a frequency of $\nu=8.64\;$GHz. This assumes a spherical outflow, 
but if the same outflow (of same $\beta_0$ and total energy $E$) is directed into a small 
fraction $f_\Omega$ of the total solid angle, this would correspond, until the deceleration 
time, to a part of a spherical flow with $E_{\rm k,iso} = E/f_\Omega$ thus increasing $t_{\rm dec}$ 
by a factor of $f_\Omega^{-1/3}$ and hardly affecting\footnote{If $F_\nu^{\rm peak}\propto E^a$ then it would change as $F_\nu^{\rm peak}(E)\to f_\Omega F_\nu^{\rm peak}(E/f_\Omega) = f_\Omega^{1-a}F_\nu^{\rm peak}(E)$, where $a=1$ for high enough radio frequencies, and it is not very far from 1 even at lower frequencies.} $F_\nu^{\rm peak}$. For lower 
frequencies the peak time can occur later, at the passage of $\nu_m$ at $\nu_a$, whichever 
occurs later. Therefore, for \grb\ the peak of this emission might be expected on a timescale 
of roughly a year or more, but it may still be detectable (perhaps even somewhat before the 
peak time) for high enough values of $n_0$, $\epsilon_B$ and $\epsilon_e$.

The emission from the sub-relativistic ejecta is expected to dominate at $t\gtrsim t_{\rm dec}$ 
over the radio afterglow from the (originally) relativistic jet if indeed its true energy is larger, 
since by that time both outflows would be Newtonian and radiate (almost) isotropically, and the more 
energetic of the two is expected to produce a higher radio flux.

\section{\grb's Location within its Host Galaxy}
\label{sec:host}

\begin{figure}
    \centering
    \includegraphics[width=0.48\textwidth,height=0.34\textwidth]{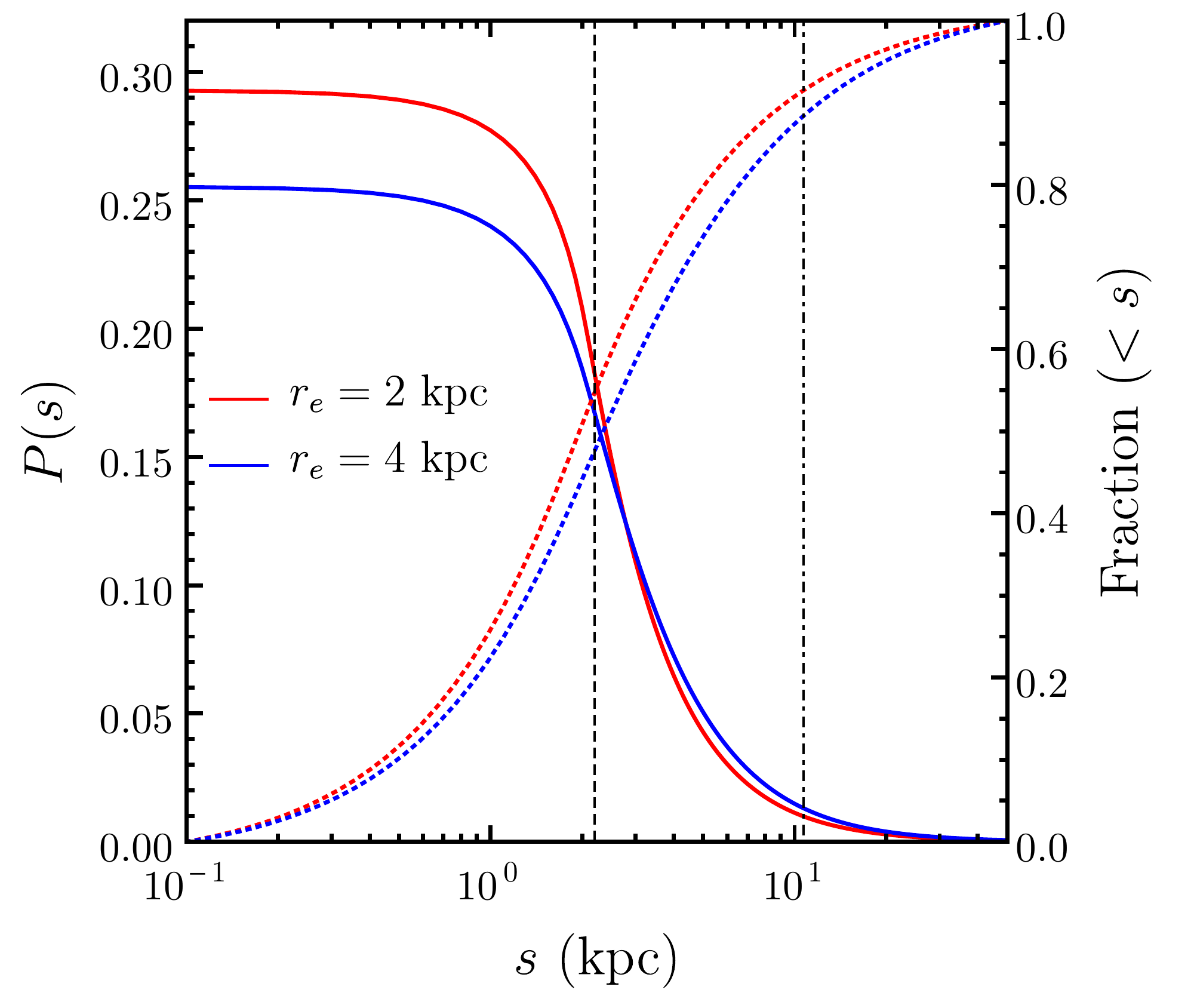}
    \caption{Probability ($P(s)$) of the binary having travelled distance $s$ from birth to merger locations 
    (solid) and the cumulative distribution
    $\int_0^s P(s')ds'$ 
    (dotted) shown for two effective radii $r_e$. The median and the distance containing 90\% of the mergers, both averaged over the 
    two cases, are shown as dashed and dot-dashed lines.}
    \label{fig:PDF}
\end{figure}

In this section, we try to determine if the position of the source in the host galaxy can be explained as a straight-line motion from the birth 
site of the binary to its merger site using a simple likelihood analysis. 
\grb\ is located $r_\perp \simeq 2\;$kpc in projection onto the plane of the sky from the center of the elliptical galaxy \host. We use $r_\perp$ to constrain the distance $s=v_{\rm sys}t_{\rm mer}$ travelled by the binary over its lifetime $t_{\rm mer}$ from its birth location $\vec{r}_i$ to that of merger, $\vec{r}_f=\vec{r}_i+\vec{s}$, assuming a straight-line motion at a systemic velocity $v_{\rm sys}$, via a likelihood 
analysis based on the stellar mass distribution in elliptical galaxies (approximated here as spherical). 
For initial and final galactocentric radii $r_i$ and $r_f=\sqrt{s^2+r_i^2-2sr_i\mu}$ where $\mu=-(\vec{r}_i\cdot \vec s)/r_is$, the probability density function of $s$ can be expressed as 
\begin{equation}\label{eq:PDF}
P(s)\propto\int_0^\infty
dr_iP(r_i)\int_{\vert r_i-s\vert}^{r_i+s}dr_fP(r_f\vert r_i,s)f_\Omega(f_f)\ .
\end{equation}
Its normalization is obtained by requiring $\int P(s)ds=1$. 

The distribution of birth locations follows that of the stellar mass in the host galaxy given here by the \citet{Hernquist90} 
profile such that $P(r_i) = d\ln{}M(r_i)/dr_i = 2r_ia/(r_i+a)^3$, where the effective radius is $r_e\approx1.8153a$. 
Given $r_i$ and $s$, the conditional probability for $r_f$ for an isotropic $P(\mu)=\frac{1}{2}$ is 
$P(r_f\vert r_i,s)=r_f/(2sr_i)$. 
Finally, given $r_f$ the fraction of observers (or total solid angle) that would see an offset $\leq r_\perp$ is $f_\Omega=1-\mu_\theta$ with $\mu_\theta=\sqrt{1-(r_\perp/r_f)^2}$ for $r_f\geq r_\perp$ and $f_\Omega=1$ for $r_f<r_\perp$. 

Fig.~\ref{fig:PDF} shows $P(s)$ along with 
the cumulative fraction of such binaries found within $s$ for two $r_e$ values bracketing the range typical for elliptical galaxies.
This suggests that the binary has traveled a rather modest distance of $s\lesssim2-10\;$kpc from its birth location. 
The distribution of merger times $t_{\rm mer}$ (including formation of the compact binary) is rather flat between 10$\;$Myr$\,$--$\,$10$\;$Gyr 
with a conspicuous peak at 20$\;$Myr due to very tight orbit binaries \citep{Belczynski+06}. 
Since in most elliptical galaxies active 
star formation ceased at $z\approx2$, the binary cannot be too young and must have $t_{\rm mer}>1\;$Gyr. Such a long-lived 
binary would in turn imply a very small systemic velocity, $v_{\rm sys} = s/t_{\rm mer}\lesssim (2 - 10)(t_{\rm mer}/1\,{\rm Gyr})^{-1}\;{\rm km~s}^{-1}$,
which is much smaller than both the expectation for a natal kick
(population synthesis studies find 
$\langle v_{\rm kick}\rangle\sim50\,$--$\,100\;{\rm km~s}^{-1}$, while larger values are often inferred from observations) and systemic velocities of stars 
in massive elliptical galaxies ($\gtrsim200\;{\rm km~s}^{-1}$). Therefore, the assumption of straight-line motion cannot hold in this case.
This can also be seen from the fact that the Keplerian times of the host galaxy's stars are typically much less than their age (i.e. many galactic orbits were completed since the last star formation epoch). One must therefore account for the host galaxy's gravitational potential,
which requires orbit-tracking numerical calculations. 
Such an effect is naturally incorporated in a population synthesis scheme which is outside the scope of this work, but see the 
recent work by \citep[][]{Abbott+17c}.


\section{Simple Analytic Model for Off-Axis Emission from GRB Jets}
\label{sec:model}
Following \citet{NPG02}, we consider an adiabatic double-sided jet with a total energy $E$ and an initial
opening angle $\theta_0$, and a simple hydrodynamic
model for the jet evolution \citep{Rhoads99,SPH99}.  Initially, as long as $\Gamma>1/\theta_0$, the jet propagates as if it were spherical with
an equivalent isotropic energy of $2E/\theta_0^2$:
\begin{equation}
E = \frac{2\pi}{3} \theta_j^2 R^3 \gamma^2 n m_p c^2 \ ,
\label{Spherical}
\end{equation}
with $\theta_j\approx\theta_0$, where $n$ is the ambient  number density. The spherical 
phase ends once $\Gamma$ drops below $1/\theta_0$, at which point  the jet expands sideways 
relativistically in it own rest frame, leading to $\theta_j\sim1/\Gamma$ and
\begin{equation}
E = \frac{2\pi}{3}  R^3  n m_p c^2 \ , \label{jet}
\end{equation}
where the shock radius remains almost constant.\footnote{More detailed calculations show 
that $R$ increases slowly and $\Gamma$ decreases exponentially with $R$ \citep{Rhoads99,Piran00}.} 
The evolution in this phase is independent of $\theta_0$, which only determines the jet 
break time, $t_j$. The lightcurves depend only on $E$ and $n$, along with the shock 
microscopic parameters (the equipartition parameters, $\epsilon_{B},\,\epsilon_{e}$, 
and the power law index $p$ of the electron distribution). During both phases the observed 
time is given by:
\begin{equation}
t=(1+z) \frac{R}{4 c \Gamma^2} \ . \label{time}
\end{equation}
Equations (\ref{jet}) and (\ref{time}) yield that the jet break
transition takes place at \citep{SPH99},
\begin{equation}
t_{j}= 0.70(1+z) \left(\frac{E_{51}}{n_0}\right)^{1/3}\left(\frac{\theta_0}{0.1}\right)^2 \, {\rm days}\ .
\end{equation}
where $Q_x$ denotes the value of the quantity $Q$ in units of
$10^x$ times its (c.g.s) units. Because of relativistic beaming, an
observer  located at $\theta_{\rm obs}$ outside the initial opening
angle of the jet ($\theta_{\rm obs}> \theta_0$) will (practically)
observe the afterglow emission only near its peak at $t_\theta$ when $\Gamma =
1/\theta_{\rm obs}$:
\begin{equation}\label{t_peak}
t_{\rm peak}(\theta_{\rm obs}) =    A \left(\frac{\theta_{\rm obs}}{\theta_0}\right)^2 t_{j} =  70(1+z)A \left(\frac{E_{51}}{n_0}\right)^{1/3}\theta_{\rm obs}^2 \, {\rm days}\ . 
\end{equation}
The flux rises until it peaks at $t_{\rm peak}$ and subsequently decays in the same way as for an on-axis observer. The
factor $A$ in Eq.~(\ref{t_peak}) is of order unity and will be taken as 1 following \citep{NPG02}. 

The synchrotron slow cooling ($\nu_m<\nu_c$) light curve for the initial
(spherical) phase was derived by \citet{SPN98} and refined by \citet{GS02}. \citet{SPH99}
provide temporal scalings for the maximal flux and the typical
synchrotron ($\nu_m$) and cooling ($\nu_c$) frequencies during the modified
hydrodynamic evolution after the jet break. Combining both
results, using the Granot \& Sari (2002) normalization for the
fluxes and typical frequencies, one obtains the universal post
jet-break light curve \citep{NPG02} for an on-axis observer, where for convenience the frequency is normalized to the optical ($5\times10^{14}\;$Hz). The flux above the self-absorption frequency is given by
\begin{eqnarray}\label{Fnu1a}
F_{\nu>\nu_c,\nu_m}(t)= 0.459\frac{g_0(p)}{g_0(2.2)}10^\frac{2.2-p}{3.93}(1+z)^\frac{p+2}{2}D_{L28}^{-2}(1+Y)^{-1} 
\\ \nonumber
\times\,\epsilon_{e,-1}^{p-1} \epsilon_{B,-2}^\frac{p-2}{4}
n_0^\frac{-p-2}{12} E_{50.7}^\frac{p+2}{3}t_{\rm days}^{-p}\nu_{14.7}^{-p/2}\
{\rm mJy} \ ,
\\ \label{Fnu1b}
F_{\nu_m<\nu<\nu_c}(t)= 0.170\frac{g_1(p)}{g_1(2.2)}10^\frac{2.2-p}{3.93}(1+z)^\frac{p+3}{2}D_{L28}^{-2}\quad\quad\quad
\\ \nonumber
\quad\times\,\epsilon_{e,-1}^{p-1} \epsilon_{B,-2}^\frac{p+1}{4}
n_0^\frac{3-p}{12}E_{50.7}^\frac{p+3}{3}
t_{\rm days}^{-p}\nu_{14.7}^{(1-p)/2}\
{\rm mJy}\ ,
\\ \label{Fnu1c}
F_{\nu_a<\nu<\nu_m<\nu_c}(t)= 3.62\frac{g_2(p)}{g_2(2.2)}(1+z)^{5/3}D_{L28}^{-2}\quad\quad\quad\quad\quad
\\ \nonumber
\times\,\epsilon_{e,-1}^{-2/3}\epsilon_{B,-2}^{1/3}n_0^{2/9}E_{50.7}^{10/9}t_{\rm days}^{-1/3}\nu_{9.93}^{1/3}\
{\rm mJy}\ ,
\end{eqnarray}
where  $D_{L}$ is the luminosity distance and $g_0(p)\equiv
10^{-0.56p}(p-0.98)\left[(p-2)/(p-1)\right]^{p-1}$, $g_1(p)\equiv
10^{-0.31p}(p-0.04)\left[(p-2)/(p-1)\right]^{p-1}$,$g_2(p)=(p-1)^{5/3}/[(3p-1)(p-2)^{2/3}]$ and $Y$ is the Compton y-parameter (which is included in the results shown in the figures below).  The cooling
frequency $\nu_c$ and the typical synchrotron frequency of the minimal energy electrons $\nu_m$ at $t>t_j$ are given by equating the flux in equation~(\ref{Fnu1b}) to that in equations (\ref{Fnu1a}) and (\ref{Fnu1c}), respectively,
\begin{eqnarray}\label{nu_c}
\nu_c = 3.62\times 10^{15}\left[\frac{2.16}{1.22}{(p-0.98)\over (p-0.04)}\right]^2
10^\frac{2.2-p}{1.985}(1+z)^{-1}\quad\quad
\\ \nonumber
\times\,\epsilon_{B,-2}^{-3/2} n_0^{-5/6}E_{50.7}^{-2/3}(1+Y)^{-2}\ {\rm
Hz}\ ,
\\ \label{nu_m}
\nu_m = 3.74\times10^{11}\left[\frac{g_1(p)g_2(2.2)}{g_1(2.2)g_2(p)}10^\frac{2.2-p}{3.93}\right]^{6/(3p-1)}(1+z)\quad\ 
\\ \nonumber
\times\,\epsilon_{B,-2}^{1/2}\epsilon_{e,-1}^{2} n_0^{-1/6}E_{50.7}^{2/3}t_{\rm days}^{-2}\ {\rm
Hz}\ .
\end{eqnarray}
Note that both frequencies are independent of $\theta_0$, and $\nu_c$ remains constant in time (at $t>t_j$).

Since the maximal flux occurs near $t_{\rm peak}(\theta_{\rm obs})$ when the off-axis lightcurve joins that for an on-axis observer, the peak flux for an observer at $\theta_{\rm obs}$ can be obtained by substituting  Eq.~(\ref{t_peak}) into Eqs.~(\ref{Fnu1a})-(\ref{Fnu1c}), 
\begin{eqnarray}\label{Fnuobs1a} 
F_{\nu>\nu_c,\nu_m}^{\rm peak}(\theta_{\rm obs}) = 1.67\frac{g_0(p)}{g_0(2.2)}A^{-p}(1+z)^{1-p/2}D_{L28}^{-2}\quad\quad\quad
\\ \nonumber
\quad\quad\quad\times\,(1+Y)^{-1}\epsilon_{e,-1}^{p-1}\epsilon_{B,-2}^\frac{p-2}{4}
n_0^\frac{3p-2}{12}E_{50.7}^{2/3}\nu_{14.7}^{-p/2}
\theta_{\rm obs,-1}^{-2p} \ {\rm mJy} \ ,
\\ \label{Fnuobs1b} 
F_{\nu_m<\nu<\nu_c}^{\rm peak}(\theta_{\rm obs}) = 0.618\frac{g_1(p)}{g_1(2.2)}A^{-p}(1+z)^{(3-p)/2}D_{L28}^{-2}\quad\ \ 
\\ \nonumber
\quad\quad\quad\quad\times\,\epsilon_{e,-1}^{p-1}\epsilon_{B,-2}^\frac{p+1}{4}
n_0^\frac{p+1}{4}E_{50.7}\nu_{14.7}^{(1-p)/2}
\theta_{\rm obs,-1}^{-2p} \ {\rm mJy} \ .
\\ \label{Fnuobs1c} 
F_{\nu_a<\nu<\nu_m<\nu_c}^{\rm peak}(\theta_{\rm obs})= 4.40\frac{g_2(p)}{g_2(2.2)}A^{-1/3}(1+z)^{4/3}D_{L28}^{-2}\quad\ \ 
\\ \nonumber
\times\,\epsilon_{e,-1}^{-2/3}\epsilon_{B,-2}^{1/3}
n_0^{1/3}E_{50.7}\nu_{9.93}^{1/3}\theta_{\rm obs,-1}^{-2/3}\
{\rm mJy}\ .
\end{eqnarray}
The peak flux (which is also independent of $\theta_0$) depends very strongly on $\theta_{\rm obs}$, and quickly decreases when the observer moves away from the jet
axis.

In this section it was implicitly assumed that the observer is initially well outside 
the edge of the jet, $\theta_{\rm obs}\gtrsim2\theta_0$. Nonetheless, it can be still 
be generalized to closer lines of sight, $\theta_0+1/\Gamma < \theta_{\rm obs} \lesssim 2\theta_0$ 
or correspondingly $1/\Gamma_0<\Delta\theta\lesssim\theta_0$ where 
$\Delta\theta\equiv\theta_{\rm obs}-\theta_0$, as follows. In this case the peak flux 
still occurs when the beaming cone reaches the line of sight, however in this regime 
it occurs before the jet break time, i.e. while $\Gamma>1/\theta_0$ and the jet has not 
yet come into lateral causal contact and therefore has not expanded sideways significantly. 
Therefore, the jet dynamics may still be approximated as part of a spherical flow with 
$E_{\rm k,iso}=E/(1-\cos\theta_0)\approx 2E/\theta_0^2$. Using the corresponding expression 
for $\Gamma(t)$ \citep[e.g.][]{SPN98,GS02,single-shell} and equating it to $1/\Delta\theta$ 
one obtains an expression for the peak time in this case,
\begin{equation}\label{eq:small_thobs}
\frac{t_{\rm peak}(1/\Gamma_0<\Delta\theta\lesssim\theta_0)}{(1+z)} \approx 1.4\times 10^4\fracb{E_{50.7}}{n_0\theta_{0,-1}^{2}}^{1/3} \fracb{\Delta\theta}{0.05}^{8/3}\;{\rm s}\ .
\end{equation}
This roughly coincides with Eq.~(\ref{t_peak}) for $\Delta\theta=\theta_0$, which corresponds to $t_{\rm peak}\sim t_j$.
The peak flux can then be obtained by substituting $t=t_{\rm peak}$ from Eq.~(\ref{eq:small_thobs}) in the expression for the on-axis pre-jet break flux density at the relevant PLS of the spectrum \citep[e.g. Table 1 of][]{GS02} corresponding to a spherical flow with $E_{\rm k,iso}\approx 2E/\theta_0^2$.

\section{General Predictions for a Nearby Short-Hard GRB}\label{sec:genres}

\begin{figure}
\centering
\includegraphics[width=0.97\columnwidth]{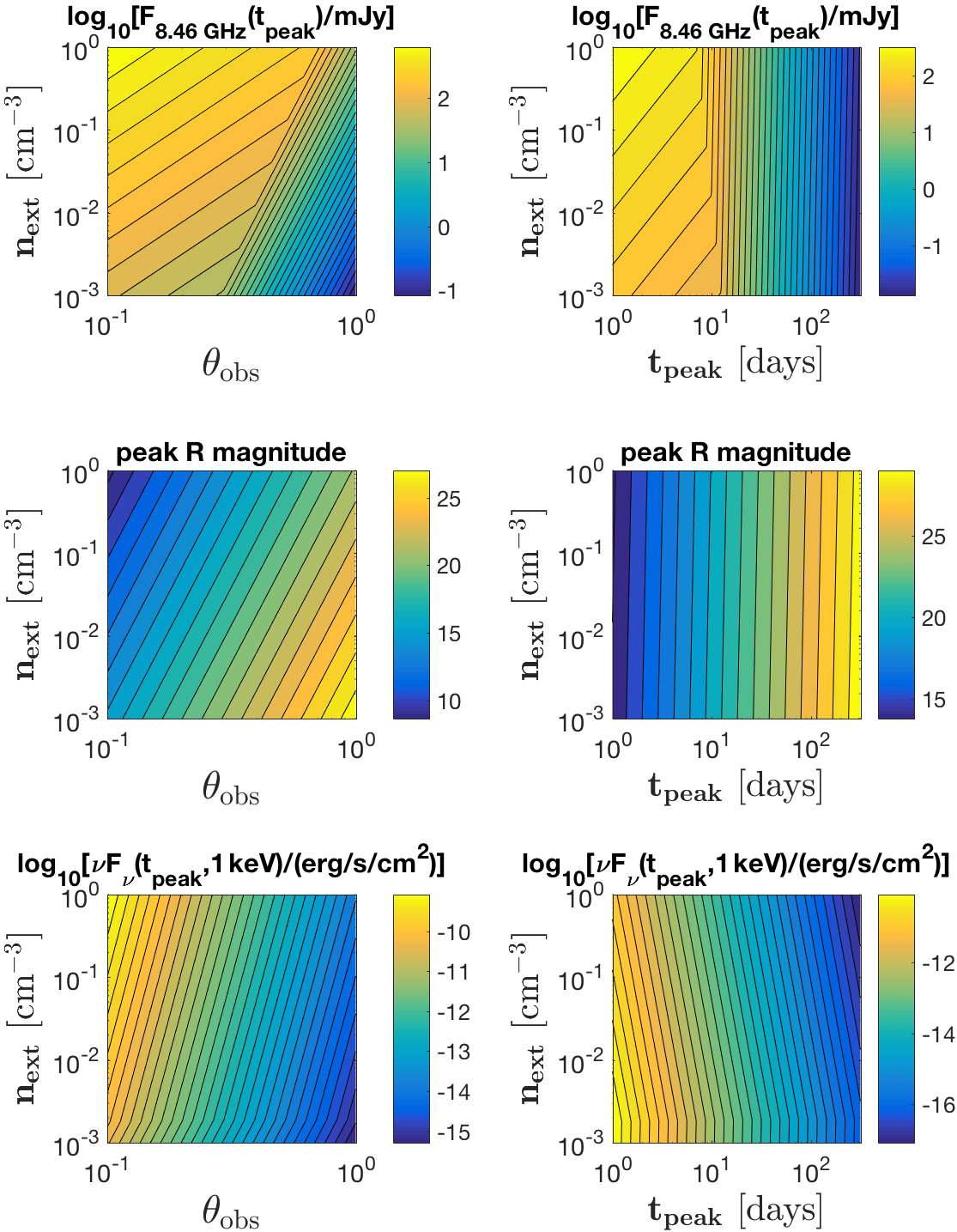} 
\caption{The peak flux in the radio ($8.46\;$GHz; {\it top panels}), optical (R-band; {\it middle panels}) and X-ray ($1\;$keV; bottom panels), for an off-axis observer as a function of the external density $n=n_0\;{\rm cm^{-3}}$ and the viewing angle $\theta_{\rm obs}$ ({\it left panels}) or the corresponding peak time $t_{\rm peak}$ ({\it right panels}). The remaining model parameters are $\epsilon_e=\epsilon_B=0.1$, $p=2.5$, $E=10^{49}\;$erg.}
\label{Contours:n}
\end{figure}

\subsection{Predictions if $\theta_{\rm obs}$ is not Known from the Gravitational Wave Signal}
From the GW observation the distance $D$ to the source is rather small and therefore we neglect here cosmological redshift and time dilation (effectively using $z=0$). For the shock microphysical parameters we choose fiducial values guided by afterglow modeling of both long and short GRBs:  $\epsilon_{e}=0.1$, $\epsilon_{B}=0.1$, $p=2.5$. For the external density we take a value
of $n_0=1$, typical of the ISM. For the jet energy we are guided by the typical isotropic equivalent gamma-ray energies of SGRB, $E_{\rm\gamma,iso}\sim10^{49}-10^{51}\;$erg, and assuming that the jet covers a fraction $\sim10^{-2}-10^{-1}$ of the total solid angle (corresponding to $8^\circ\lesssim\theta_0\lesssim26^\circ$), or $E\sim10^{48}-10^{49}\;$erg, and take the upper end of this estimated range, $E=10^{49}\;$erg. For illustrative purposes we assume a distance to the source of $D=40\;$Mpc. We explore the dependence of the peak flux on the most uncertain parameters, namely the external density $n$, the viewing angle $\theta_{\rm obs}$, the fraction of the internal energy behind the afterglow shock in the magnetic field, $\epsilon_B$, and the jet energy, $E$. Semi-analytic modeling of 
lightcurves for different viewing angles and from different emission components, namely prompt, afterglow, and the hot cocoon surrounding the relativistic jet, was also done in \citet{Lazzati+17a}.

Fig.~\ref{Contours:n} shows the dependence of the peak flux in X-ray ($1\;$keV), optical (R-band) and radio ($8.46\;$GHz) on the external density $n$ and the viewing angle $\theta_{\rm obs}$ as well as on the corresponding peak time, $t_{\rm peak}$.

\begin{figure}
\centering
\includegraphics[width=0.97\columnwidth]{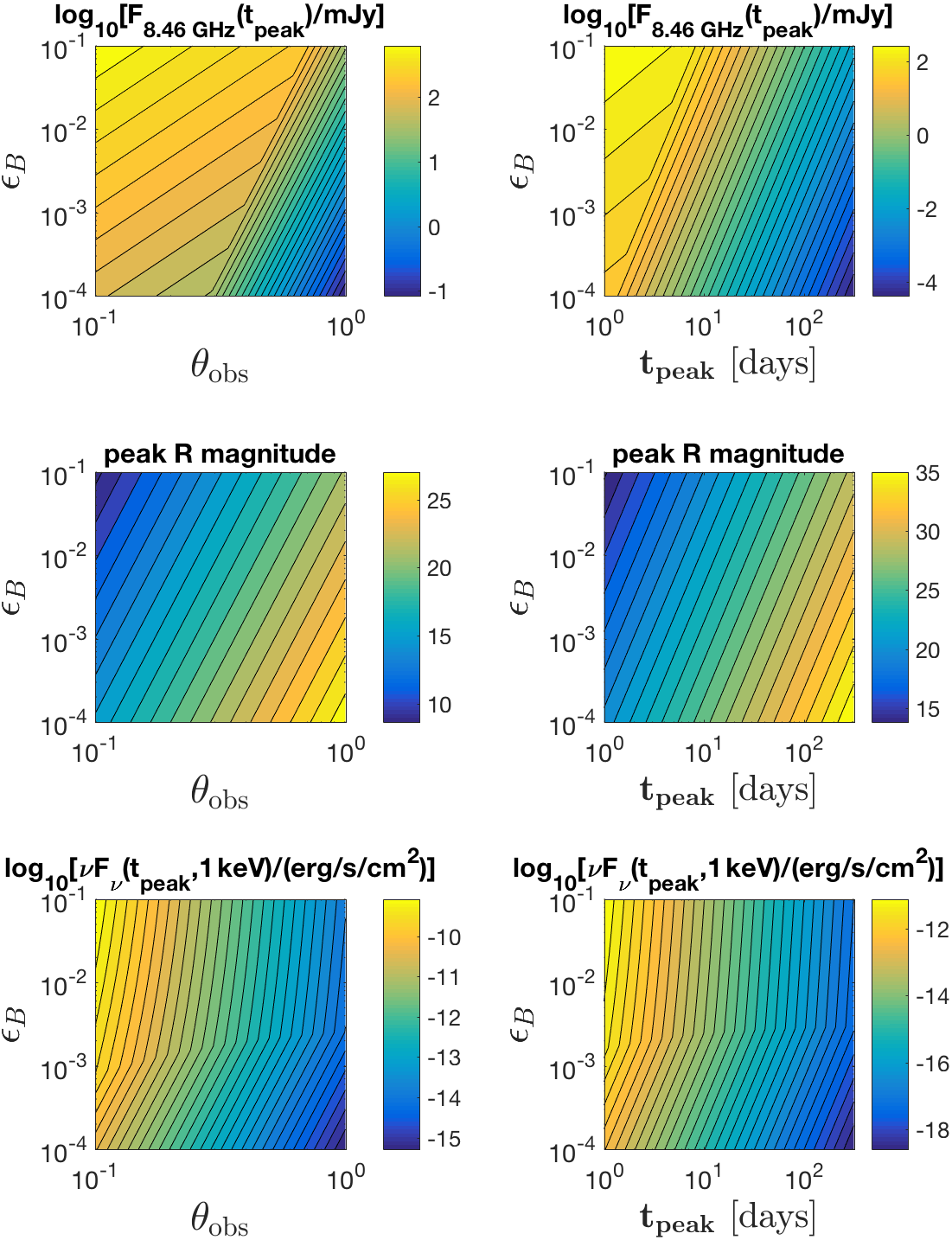}
\caption{  Similar to Fig.~\ref{Contours:n} but varying the magnetic field equipartition parameter $\epsilon_B$ instead of the external density (with $n_0=1$, $\epsilon_e=0.1$, $p=2.5$, $E=10^{49}\;$erg).}
\label{Contours:eb}
\end{figure}

Fig.~\ref{Contours:eb} shows the dependence of the peak flux X-ray, optical and radio on the magnetic field equipartition parameter $\epsilon_B$, and on the viewing angle $\theta_{\rm obs}$ or the corresponding peak time, $t_{\rm peak}$.

\begin{figure}
\centering
\includegraphics[width=0.97\columnwidth]{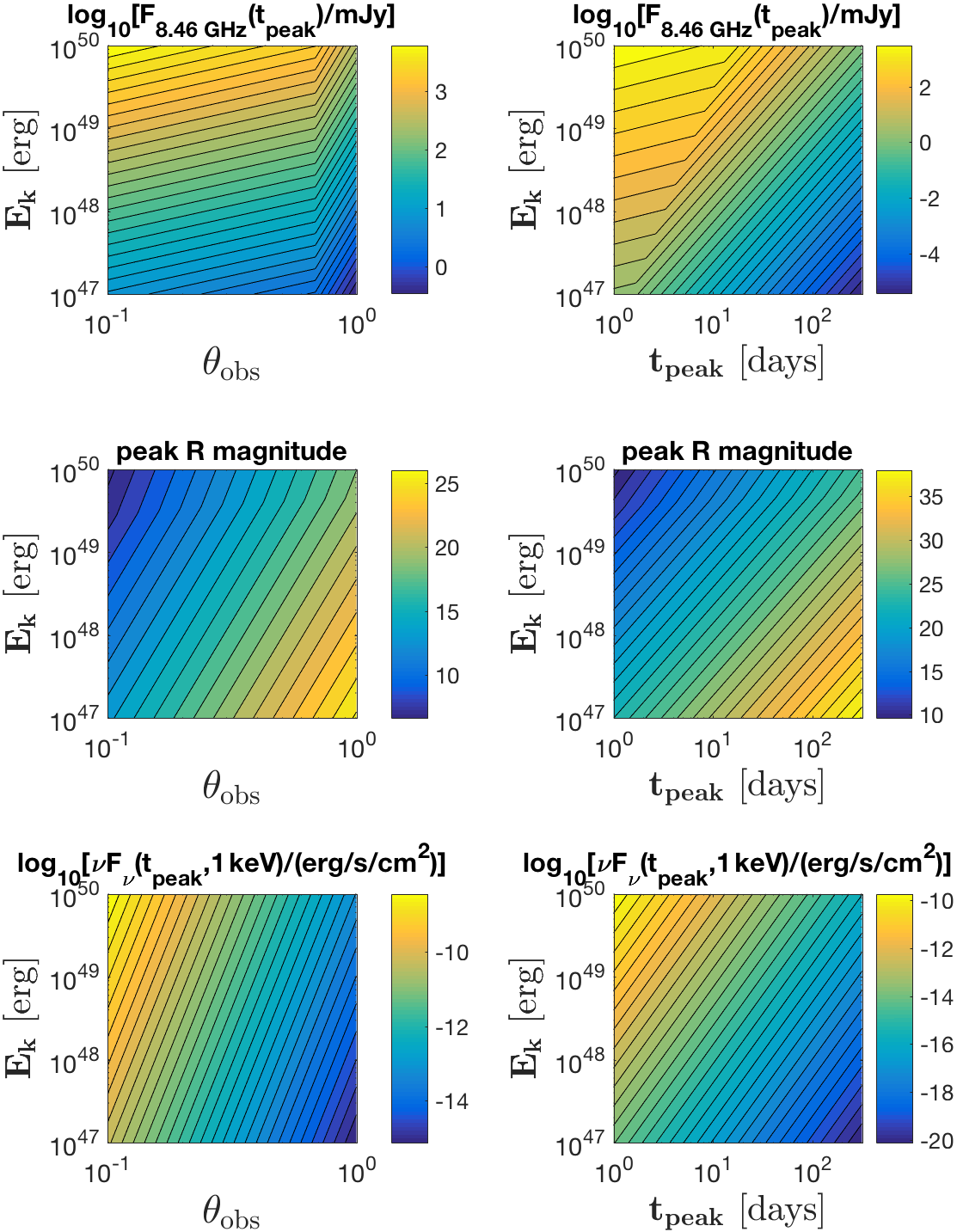} 
\caption{ Similar to Fig.~\ref{Contours:n} but varying the jet energy $E$ instead of the external density (with $n_0=1$, $\epsilon_e=\epsilon_B=0.1$, $p=2.5$).}
\label{Contours:E}
\end{figure}

Fig.~\ref{Contours:E} shows the dependence of the peak flux X-ray, optical and radio on the jet energy, $E$, and on the viewing angle $\theta_{\rm obs}$ or the corresponding peak time, $t_{\rm peak}$.

\subsection{More Constraints if $\theta_{\rm obs}$ is known from the Gravitational Wave Signal}
In this subsection we demonstrate how the constraints on the relevant model parameters can become tighter if the viewing angle $\theta_{\rm obs}$ is determined from the GW signal.
The latter may be achieved since the jet axis is expected to be aligned with the spin axis of the BH that is produced during the merger (since both are expected to be aligned with the binary's orbital angular momentum, i.e. normal to the orbital plane), and one can determine from the GW signal the angle of the BH's spin relative to our line of sight, which is identified with $\theta_{\rm obs}$. For concreteness, we will assume values of $\theta_{\rm obs}=0.6\pm0.1$ as a case study.

\begin{figure}
\centering
\includegraphics[width=0.97\columnwidth]{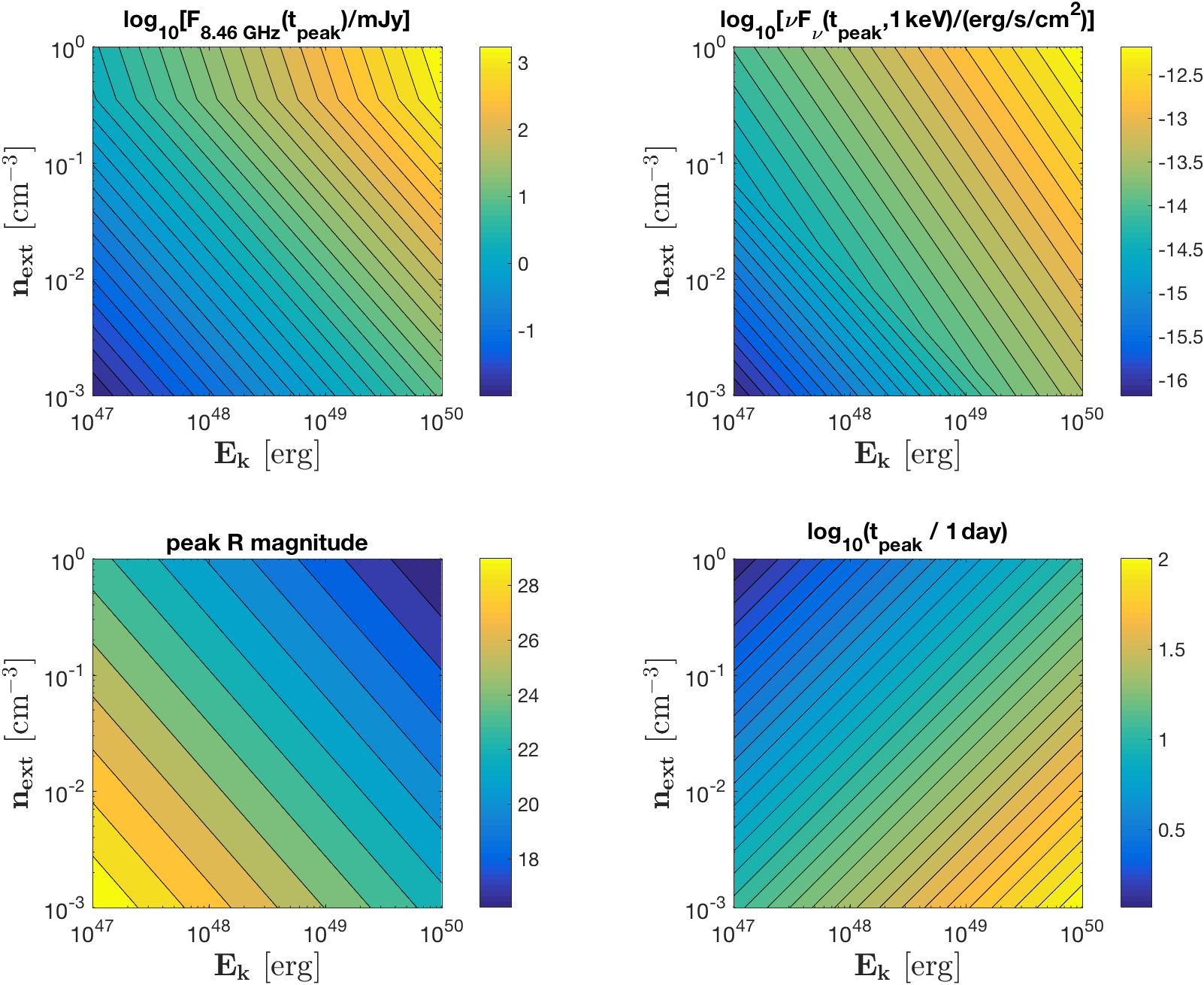} \caption{
The peak flux in the radio ($8.46\;$GHz; {\it top panels}), optical (R-band; {\it middle panels}) and X-ray ($1\;$keV; bottom panels), as well as the time of the peak flux $t_{\rm peak}$, for an off-axis observer at $\theta_{\rm obs}=0.6$, as a function of the external density $n=n_0\;{\rm cm^{-3}}$ and the jet energy $E$.}
\label{fig:Contours:n_E} 
\end{figure}

\begin{figure}
\centering
\includegraphics[width=0.97\columnwidth]{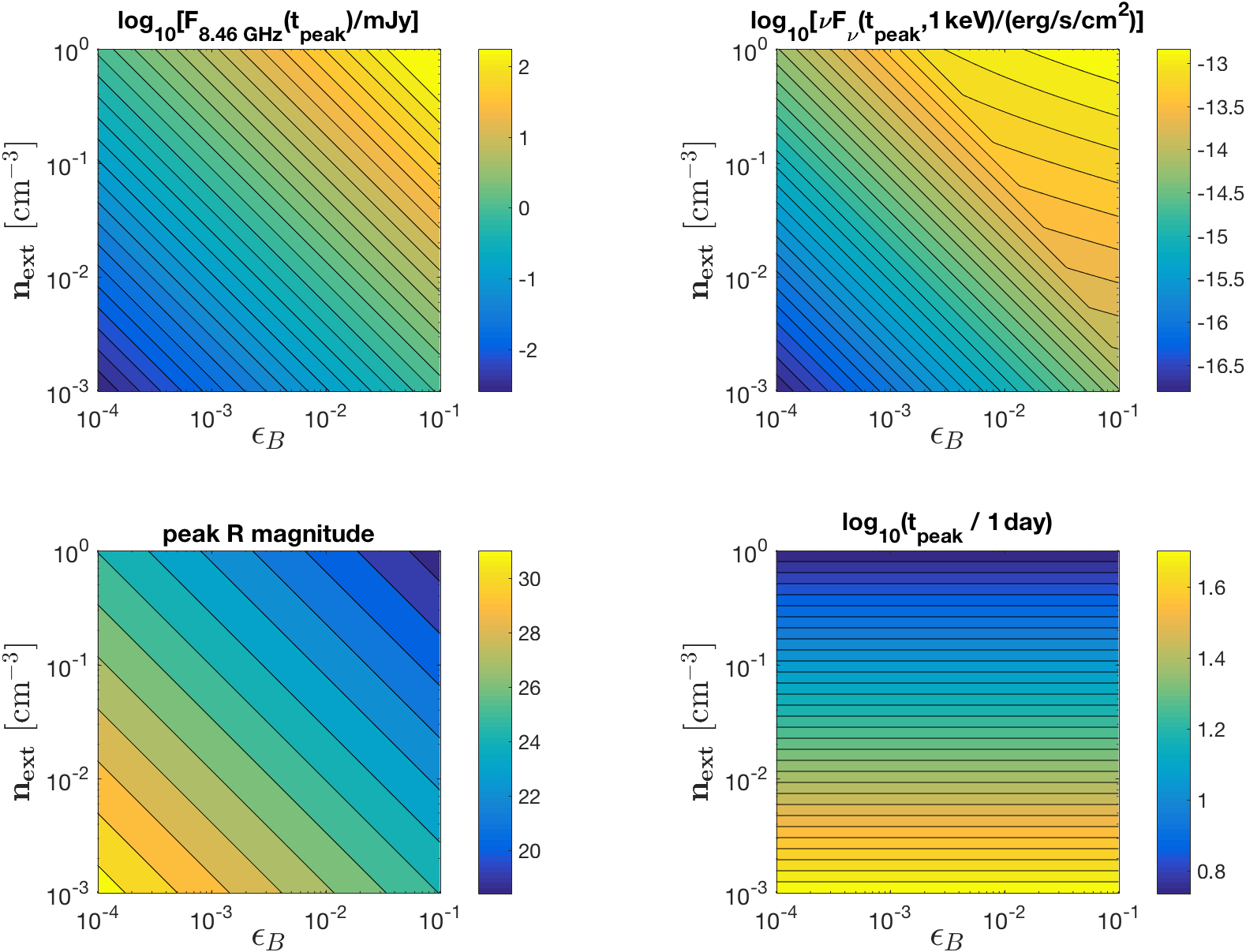} \caption{ 
The peak flux in the radio ($8.46\;$GHz; {\it top panels}), optical (R-band; {\it middle panels}) and X-ray ($1\;$keV; bottom panels), as well as the time of the peak flux $t_{\rm peak}$, for an off-axis observer at $\theta_{\rm obs}=0.6$, as a function of the external density $n=n_0\;{\rm cm^{-3}}$ and the magnetic field equipartition parameter $\epsilon_B$.}
\label{fig:Contours:n_epsB}
\end{figure}

When $\theta_{\rm obs}$ is measured from the GW signal its value can be used, which reduces the unknown parameter space and allows us to plot contour plots for two intrinsic model parameters, such as $n$ and $E$ as demonstrated in Fig.~\ref{fig:Contours:n_E}, or $n$ and $\epsilon_B$ as demonstrated in Fig.~\ref{fig:Contours:n_epsB}. Larger values of $E$, $n$ and $\epsilon_B$ would make the emission from such an off-axis GRB afterglow jet brighter at the peak time of the lightcurve, $t_{\rm peak}$, and therefore easier to detect.
On the other hand, sufficiently low values of these parameters might make such an emission too dim to be detected. The peak time scales as $t_{\rm peak}\propto(E/n)^{1/3}\theta_{\rm obs}^2$ (see Eq.~(\ref{t_peak})), and therefore even for a given $\theta_{\rm obs}$ it can vary over a reasonable range for different values of $E/n$ (see the bottom right panel of Fig.~\ref{fig:Contours:n_E}).

\section{Off-Axis Afterglow Lightcurves from GRB Jet Simulations}
\label{sec:sims}
The GRB afterglow lightcurves, and in particular those for off-axis observers ($\theta_{\rm obs}>\theta_0$), 
strongly depend on the jet dynamics during the afterglow phase. Analytic models have traditionally obtained 
an exponential lateral expansion with radius after the jet's Lorentz factor $\Gamma$ drops below the inverse 
of its initial half-opening angle $\theta_0$, $1\ll\Gamma<1/\theta_0$ \citep[e.g.,][]{Rhoads99,SPH99,Gruzinov07,KK15}. 
However, numerical simulations \citep[e.g.,][]{Granot01,ZM09} have found that the jet's lateral expansion 
is much more modest and most of its energy remains within its initial half-opening angle until it becomes 
mildly relativistic, and only then its starts to gradually approach spherical symmetry (i.e. the Newtonian 
spherical self-similar Sedov-Taylor solution). 

This apparent discrepancy has been reconciled \citep{WWF11,GP12} by showing that the simple analytic models 
strongly rely on the approximations of a small jet half-opening angle ($\theta_j\ll1$) and ultra-relativistic 
Lorentz factor ($\Gamma\gg 1$) and break down when the jet is no longer extremely narrow and ultra-relativistic. 
Rapid, exponential lateral expansion with radius is expected only for jets that are initially extremely 
narrow, $\theta_0\ll10^{-1.5}$ \citep{GP12}. Inferred values of $\theta_0$ in GRBs are typically not that 
small, and therefore a more modest lateral expansion is expected, as seen in numerical simulations. 

Moreover, simulations also show that the jet is not uniform as assumed for simplicity in analytic models. 
Instead, at the front part of the jet near its head is the fastest and most energetic fluid whose velocity 
is almost in the radial direction, while at the sides of the jet there is slower and less energetic fluid 
whose velocity is not in the radial direction but points more sideways.\footnote{The direction velocity 
of the fluid just behind the shock is in the direction of the local shock normal in the rest frame of the 
upstream fluid, i.e. that of the circumstellar medium and central source \citep[from the shock jump conditions, as pointed out, e.g., by][]{KG03}.} This slower material at the sides of the jet dominates the emission at early times for 
large viewing angles $\theta_{\rm obs}$, since its emission has a much wider beaming cone and covers a larger 
solid angle, as compared to the material at the front of the jet, which has a much larger Lorentz factor $\Gamma$ 
and its velocity is almost in the radial direction, so that its radiation is strongly beamed away from 
observers at large off-axis viewing angles \citep{Granot01,off-axis2002,Granot07,ZM09,DeColle12b,vEM11}.

\begin{figure}
\centering
\includegraphics[width=0.97\columnwidth]{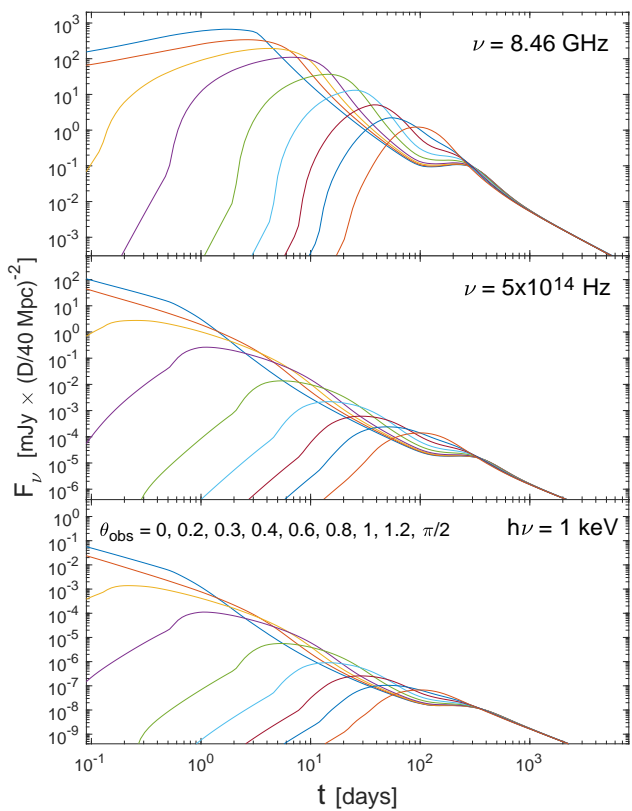} 
\caption{Afterglow lightcurves from numerical simulations for different viewing angles $\theta_{\rm obs}$ from the jet's symmetry axis, in the radio ({\it upper panel}), optical ({\it middle panel}) and X-ray ({\it bottom panel}). The simulations follow \citet{DeColle12a,DeColle12b} and are for an external density $n_0=1$, true (double-sided) jet energy of $E=10^{49}\;$erg, initial jet half-opening angle of $\theta_0=0.2$, as well as $\epsilon_e=\epsilon_B=0.1$ and $p=2.5$. The flux normalization is for a distance of $D=40\;$Mpc to the source.}
\label{fig:LC_off-axis}
\end{figure}

Figure~\ref{fig:LC_off-axis} shows afterglow lightcurves from 2D relativistic hydrodynamic simulations of a GRB jet following \citet{DeColle12a,DeColle12b}. 
The initial conditions feature a conical wedge of initial half-opening angle $\theta_0=0.2$ from the \citet{BM76} self-similar solution, 
with a true energy (for a double-sided jet) of $E=10^{49}\;$erg (corresponding to an isotropic equivalent kinetic energy of $E_{\rm k,iso}\approx5\times10^{50}\;$erg),
and an external density of $n=1\;{\rm cm^{-3}}$ (i.e. $n_0=1$).
The lightcurves for observers located at different viewing angles $\theta_{\rm obs}$ from the jet's 
symmetry axis are calculated following the hydrodynamic simulation of the jet dynamics, where we have 
set $\epsilon_e=\epsilon_B=0.1$ and $p=2.5$ for the values of the shock microphysical parameters.

It can be seen that the lightcurves for larger viewing angles $\theta_{\rm obs}$ peak at a later time and at a lower flux level. For a given $\theta_{\rm obs}$, after the lightcurve peaks it approaches that for an on-axis observer, since the jet's beaming cone engulfs the line of sight. At very later times when the flow becomes Newtonian the lightcurves become essentially independent 
of the viewing angle $\theta_{\rm obs}$ since relativistic beaming and light travel effects become 
unimportant. Around the non-relativistic transition time there is a bump in the lightcurve as the 
emission from the counter-jet becomes visible and at its peak it is somewhat brighter than the jet 
that points more towards us since relativistic beaming becomes small and due to light travel effects 
we are seeing its emission from a smaller radius (compared to that of the forward jet at the same 
observed time) where it was intrinsically brighter.

\begin{figure}
\centering
\includegraphics[width=0.97\columnwidth]{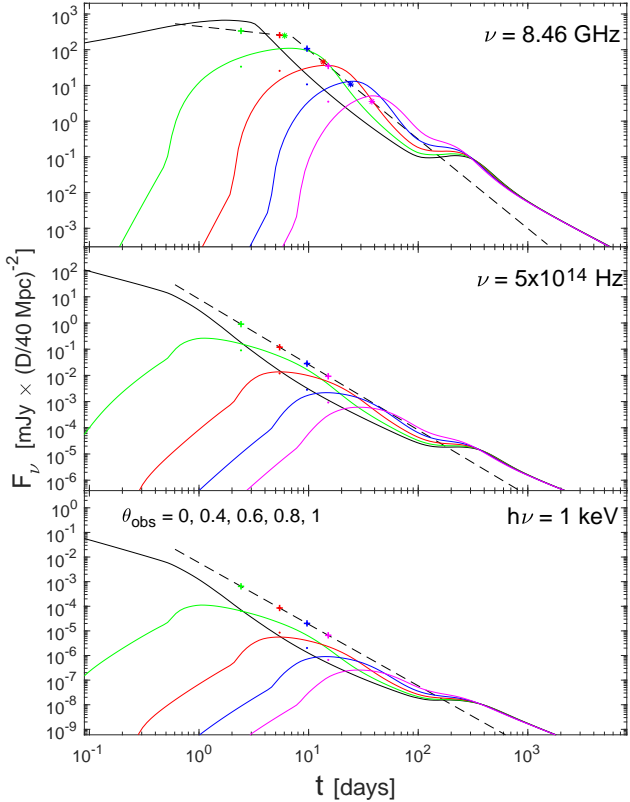} 
\caption{Similar to Fig.~\ref{fig:LC_off-axis} but showing fewer viewing angles and superimposing the analytic predictions for the peak time and flux as describes in \S\S~\ref{sec:model} and \ref{sec:genres}. The plus signs are according to Eq.~\ref{t_peak} with $A=1$ and are along the dashed black line indicating the analytic on-axis flux. The dots are a factor of 10 lower flux. For the radio ({\it top panel}) we also show asteriscs (with the same color coding for the different viewing angles $\theta_{\rm obs}$) corresponding to $A=2.5$.}
\label{fig:LC_comp_off-axis}
\end{figure}

Figure~\ref{fig:LC_comp_off-axis} compares afterglow lightcurves from numerical simulations to the analytic predictions for the peak time and flux as describes in \S\S~\ref{sec:model} and \ref{sec:genres}. In the optical and X-ray (where the on-axis lightcurves decay at early times) the lightcurves from viewing angles slightly outside of the jet $\theta_0<\theta_{\rm obs}\lesssim(2-3)\theta_0$ peak earlier than the analytic predictions due to some lateral spreading of the jet and the slower material on its sides that both cause the jet's beaming cone to reach such viewing angles faster than the analytic expectations. For larger viewing angles, $\theta_0\sim0.8-1$ in our case, the lightcurves peak later than the analytic expectation since the jet expands sideways and decelerates much more slowly with radius and observed time compared to the analytic models that feature an exponential lateral expansion, and therefore its beaming cone reaches large $\theta_{\rm obs}$ at later times. The analytic peak flux prediction is higher than that from numerical simulations by roughly an order of magnitude. The numerical flux at the analytic peak time is even somewhat below the numerical peak flux (sometimes by more than an order of magnitude; see the dots in Fig.~\ref{fig:LC_comp_off-axis} that are a factor of 10 lower flux than the analytic peak flux) since the numerical peak time generally deviates from the analytic prediction as discussed above.

In the radio the peak of the off-axis lightcurves occurs significantly later.  At the time when the line of sight enters the jet's beaming cone there is a flattening in the lightcurve but since the on-axis lightcurve still rises in the radio then unlike in the optical and X-ray where this is enough for the flux to start decaying at that time (so that it corresponds to the peak time), in the radio the flux continues to gently rise. The radio flux peaks and starts to decay only around the time when the break frequency $\nu_m$ crosses the observed radio frequency (which is observed at somewhat different times for different viewing angles $\theta_{\rm obs}$). 
In this case using $A=2.5$ gives a better fit for the peak time and flux (compared to $A=1$).

\section{Constraining the Jet Physical Parameters from \grb\ Data}
\label{sec:compdata}

\begin{figure*}
    \centering
    \includegraphics[width=0.8\textwidth]{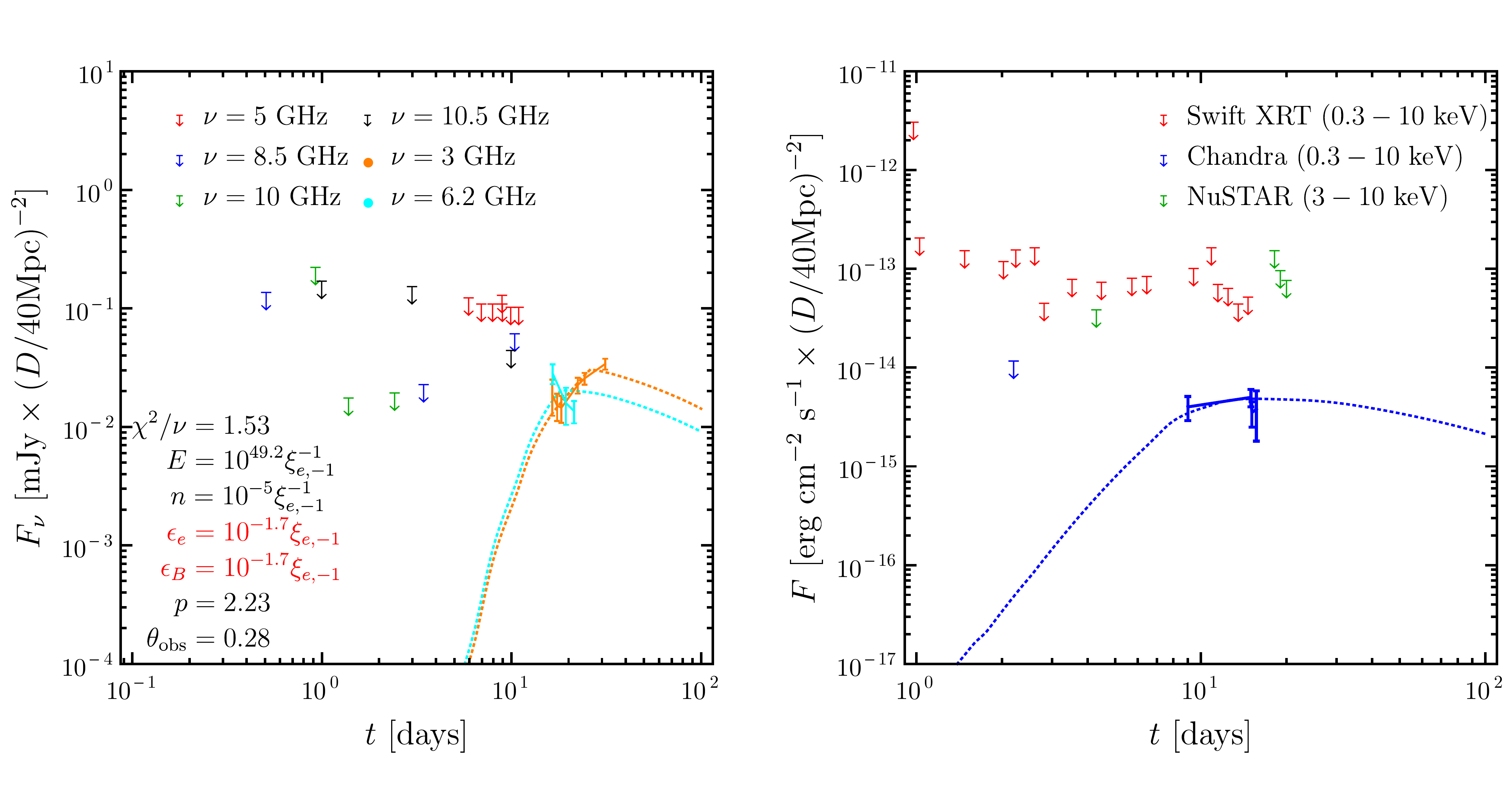}
    \includegraphics[width=0.8\textwidth]{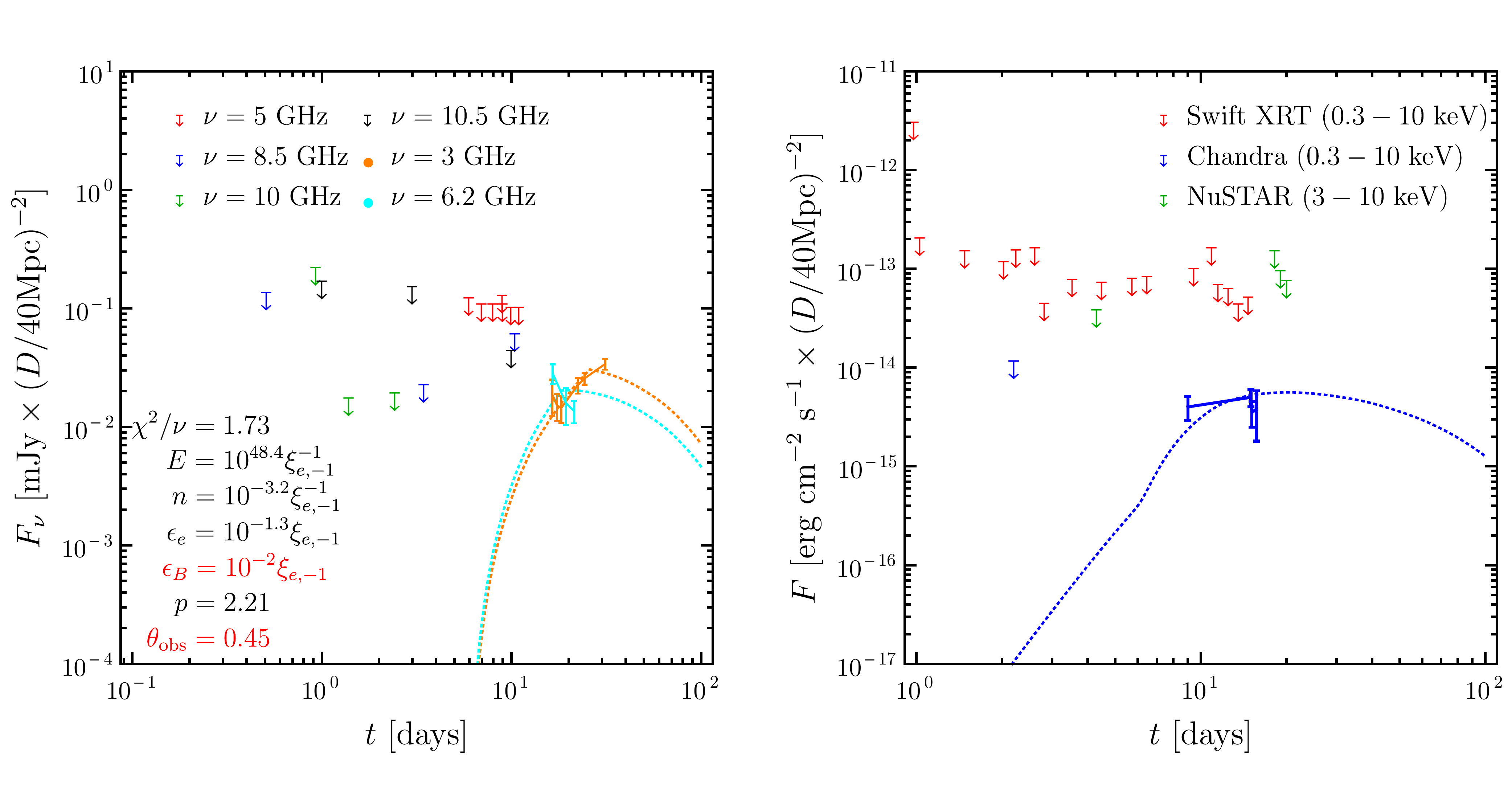}
    \includegraphics[width=0.8\textwidth]{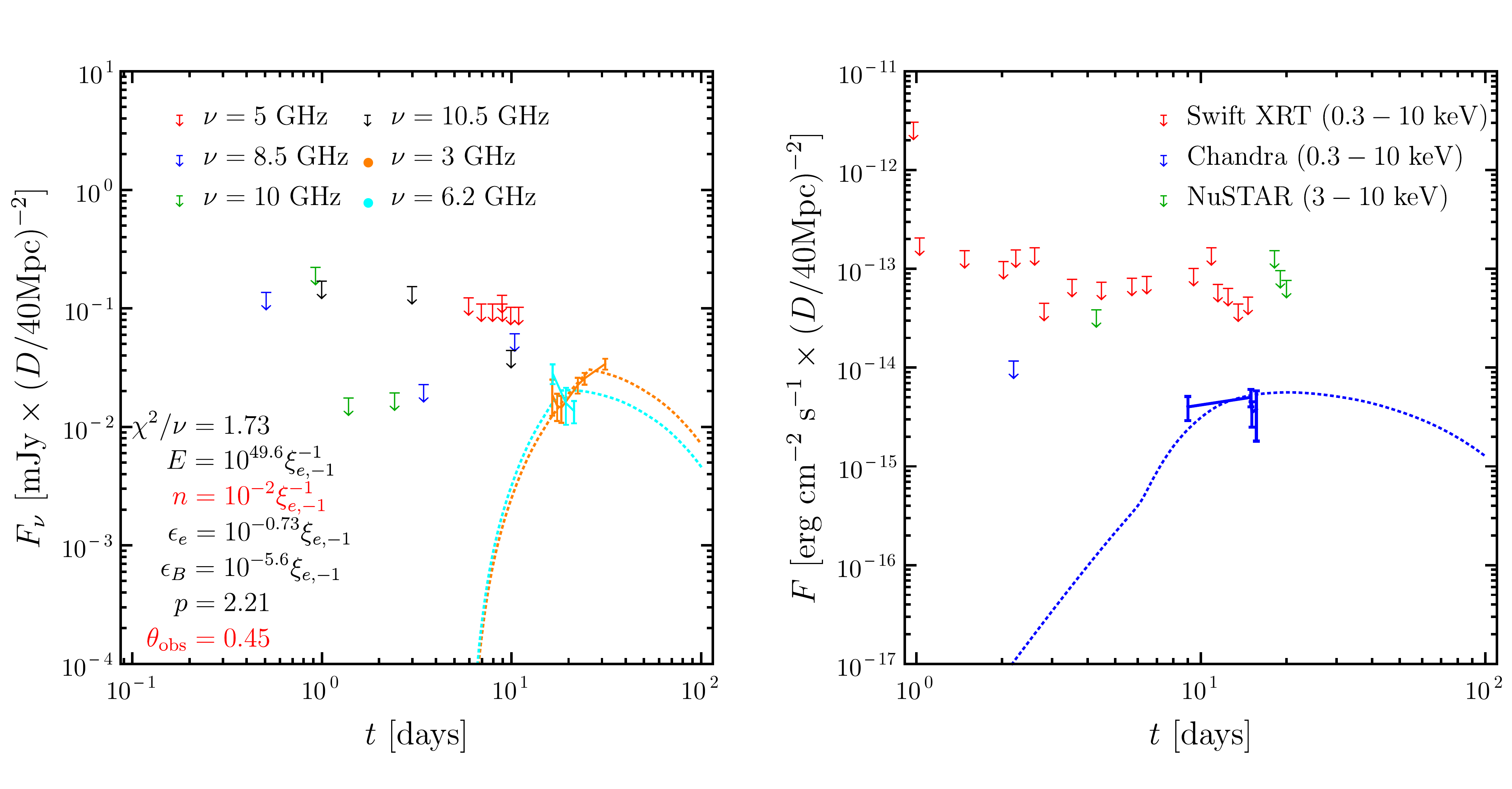}
    \caption{Least-squares afterglow model fits (dotted line) to the radio and X-ray detections (shown as dots). 
    The upper limits are shown as downward arrows. The afterglow model is described by six parameters from which 
    we fix two (shown in red), namely $\epsilon_e$ and $\epsilon_B$ (top), or $\epsilon_B$ and $\theta_{\rm obs}$ 
    (middle), or $\theta_{\rm obs}$ and $n$ (bottom), while fitting for the remaining four. The reduced chi-square 
    $\chi^2/\nu$ is comparable in all three fits. The fraction of electrons that are accelerated to a non-thermal power-law energy distribution and contribute 
    to the flux are assumed to be $\xi_e=10^{-1}\xi_{e,-1}$.}
    \label{fig:par-plt}
\end{figure*}

\citet{short} have argued that 
the relatively low measured values of the isotropic equivalent $\gamma$-ray energy output, $E_{\rm\gamma,iso}=(5.36\pm0.38)\times10^{46}D_{40\,\rm{Mpc}}^2\;$erg, and the peak $\nu F_\nu$ photon energy,  $E_{\rm p}\sim40-185\;$keV, measured for \grb\ favor an off-axis viewing angle outside of the jet's initial aperture ($\theta_{\rm obs}>\theta_0$). Comparison of these values to those typical of other short-hard GRBs (that are viewed from within the jet's initial aperture ($\theta_{\rm obs}<\theta_0$) implies that for a uniform sharp-edged jet our line of sight is only slightly outside of the jet $\Delta\theta = \theta_{\rm obs}-\theta_0\sim (0.05-0.1)(100/\Gamma_0)$, which would in turn imply an unusually high on-axis $E_{\rm p}\sim3-8\;$MeV (or $\sim5-20\;$MeV for the main half-second initial spike). Such a viewing angle would also imply an early peak for the afterglow lightcurve, since the beaming cone would reach the line of sight within $t_{\rm peak}\lesssim 1\;$day, when $\Gamma(t_{\rm peak})\approx 1/\Delta\theta\sim (10-20)(\Gamma_0/100)$ (see Eq.~(\ref{eq:small_thobs})). Since such an early peak of the afterglow emission was not seen in \grb, this scenario is disfavored.    

An alternative scenario that was favored by \citet{short} is that the jet does not have sharp edges but instead a roughly uniform core of half-opening angle $\theta_0$ outside of which the energy per solid angle drops gradually, rather than abruptly.
In this picture the prompt GRB is dominated by the emission from material along our line of sight, which is well outside of the jet's core angle ($\theta_{\rm obs}\gtrsim2\theta_0$) and correspondingly less energetic. This would imply a higher afterglow flux at early times compared to a sharp-edged jet, arising from the material along our line of sight. The latter can be estimated by assuming spherical emission with the local isotropic equivalent kinetic energy \citep[e.g.,][]{GS02}, which is expected to be comparable to that observed in gamma-rays,
$E_{\rm k,iso}\sim E_{\rm\gamma,iso}\approx5.4\times10^{46}D_{40\,\rm{Mpc}}^2\;$erg. The resulting flux densities in the relevant power-law segments of the spectrum are \citep{GS02}:
\begin{eqnarray}\label{eq:Fnu3a}
F_{\nu>\nu_c,\nu_m}(t)= 1.18\frac{g_0(p)}{g_0(2.2)}10^{1.78(2.2-p)}D_{40\,{\rm Mpc}}^{-2}(1+Y)^{-1}\quad
\\ \nonumber
\times\,\epsilon_{e,-1}^{p-1} \epsilon_{B,-2}^{p-2\over 4}
E_{46.64}^{p+2\over 4}
t_{\rm days}^{(2-3p)/4}\nu_{14.7}^{-p/2}\
\mu{\rm Jy} \ ,\quad 
\\ \label{eq:Fnu3b}
F_{\nu_m<\nu<\nu_c}(t)= 0.0131\frac{g_1(p)}{g_1(2.2)}10^{1.78(2.2-p)}D_{40\,{\rm Mpc}}^{-2}\quad\quad\quad\ 
\\ \nonumber
\times\,\epsilon_{e,-1}^{p-1} \epsilon_{B,-2}^{p+1\over 4}
n_0^{1/2} E_{46.64}^{p+3\over 4}
t_{\rm days}^{3(1-p)/4}\nu_{14.7}^{(1-p)/2}\
\mu{\rm Jy}\ ,
\\ \label{eq:Fnu3c}
F_{\nu_a<\nu<\nu_m<\nu_c}(t)= 0.196\frac{g_2(p)}{g_2(2.2)}D_{40\,{\rm Mpc}}^{-2}\quad\quad\quad\quad\quad\quad\quad\ \ 
\\ \nonumber
\times\,\epsilon_{e,-1}^{-2/3}\epsilon_{B,-2}^{1/3}
n_0^{1/2}E_{46.64}^{5/6}
t_{\rm days}^{1/2}\nu_{9.93}^{1/3}\
{\rm mJy}\ .
\end{eqnarray}

In the remainder of this section we compare the predicted afterglow emission from a GRB jet viewed 
off-axis to the observations of \grb. For early emission, before the peak of the off-axis lightcurve 
$t_{\rm peak}$ that occurs when the beaming cone from the jet's core reaches our line of sight, we 
use the above analytic expressions for the emission from the material along our line of sight. In 
addition, we calculate the flux from the jet's core and the wings that is produced as a result of 
its interaction with the external medium by using the results of hydrodynamic simulations for the 
jet dynamics during the afterglow phase \citep{DeColle12a,DeColle12b}. These simulations use as 
initial conditions a conical wedge taken out of the spherical self-similar \citet{BM76} solution. 
This is the sharpest-edged jet one can take (with a step function at $\theta=\theta_0$), but still 
on the dynamical time it develops wings of slower and less energetic material on its sides, at large 
angles $\theta$, whose emission is less beamed than that from its energetic core (at $\theta<\theta_0$).

The numerical lightcurves are calculated by post-processing the results of the hydrodynamic simulation. However, calculating the lightcurves for a very large number of sets of parameter values would require huge computational resources. This is a serious issue not only when varying all of the free parameters ($E$, $n$, $\epsilon_e$, $\epsilon_B$, $p$, $\theta_{\rm obs}$, and $\theta_{0}$), but even when varying only a subset of them. We address this issue as follows.
First, we use the results of a single hydrodynamic simulation for $\theta_0=0.2$, $n_0=1$, and $E_{\rm k,iso}=10^{53}\;$erg corresponding to $E=(1-\cos\theta_0)E_{\rm k,iso}\approx E_{\rm k,iso}\theta_0^2/2 = 2\times10^{51}\;$erg, and rescale them to arbitrary values of $n_0$ and $E$ using the scaling relations described in \citet{scaling}. Since $\theta_0$ cannot be rescaled it remains fixed at $\theta_0=0.2$.
Second, we also use the fully analytic scaling  of the flux density within any given power-law segment (PLS) of the afterglow synchrotron spectrum with all of the model parameters, as summarized in Table~2 of \citet{scaling}. 
A broadly similar approach was used, e.g., by \citet{vanEerten-MacFadyen-2012} and \citet{Ryan+15}.

We have calculated the lightcurves at a single observed frequency, which was assumed in turn to always be 
within one of the relevant PLSs \citep[PLSs D, G and H using the notations of][]{GS02}, where each PLS is 
fully rescalabe as discussed above. For each PLS we calculated the lightcurves for a large number of 
viewing angles in the range $\theta_{\rm obs}\in[0,\pi/2]$, and then interpolated between these values for 
any arbitrary viewing angle. Since the spectrum is convex, a broken power-law approximation of the spectrum 
is simply obtained by using the minimal flux out of that for all of the relevant PLSs. The spectrum can be 
refined to a more realistic shape that is smooth near the break frequencies using, e.g., the prescriptions 
from \citet{GS02}. We find that this gives comparably good fits in terms of their $\chi^2/\nu$ but does not 
qualitatively reproduce the different temporal behavior of the two radio frequencies (since the peak in the 
lightcurve due to the passage of $\nu_m$ stretches over a much longer $\Delta\log t$). 

Our aim is to constrain our model parameters: the jet's true energy $E$, the external density $n=n_0\;{\rm cm^{-3}}$, our viewing angle $\theta_{\rm obs}$, 
and the shock microphysical parameters $\epsilon_e$, $\epsilon_B$, and $p$. To this end, we make use of the entire X-ray and radio data upper-limits given 
in \citet[][]{capstone} as well as the X-ray detections by \textit{Chandra} \citep[][]{Troja+17,Haggard+17} and radio detections by the Very Large Array 
\citep[VLA;][]{Hallinan+17}. In our case the data are not constraining enough to uniquely determine the values of all the model parameters. Therefore, we 
instead provide a few illustrative fits.

Moreover, even for ideal data there is a degeneracy that still remains. As shown by \citet{EW05}, the afterglow flux is invariant under the change $E\to E/\xi_e$, 
$n\to n/\xi_e$, $\epsilon_e\to\epsilon_e\xi_e$, and $\epsilon_B\to\epsilon_B\xi_e$, for $m_e/m_p<\xi_e\leq1$, 
where $\xi_e$ is the fraction of electrons that are shock accelerated to the relativistic power-law energy 
distribution considered in most works. This degeneracy can only be broken through the effect of the remaining 
electrons that are typically expected to form a thermal distribution \citep[e.g.][]{EW05,GS09,RL17}. Here we 
account for this degeneracy by providing the values of $E$, $n$, $\epsilon_e$ and $\epsilon_B$ as a function of $\xi_e$.

Fig.~\ref{fig:par-plt} shows three illustrative fits obtained using a non-linear least-squares method, in which we have fixed the 
values of two model parameters, for e.g. $\epsilon_e$ and $\epsilon_B$ (top-panel), or $\epsilon_B$ and $\theta_{\rm obs}$ 
(middle-panel), or $\theta_{\rm obs}$ and $n$ (bottom-panel), and allowed four parameters to vary. Each fit is shown in a separate 
row with the radio and X-ray lightcurves along with the data and the corresponding model parameter values and the reduced chi-square, 
$\chi^2/\nu$. In fixing $\theta_{\rm obs}=0.45$ we are guided by the upper limit of $\theta_{\rm obs}<\theta_{\rm obs,max}\approx0.49$ 
from the GW observation \citep{Abbott+17b}. The other parameters are fixed based on their typical values that are inferred from other 
SGRBs. Given the dearth of data in both radio and X-rays, we obtain a satisfactory $\chi^2/\nu$ for $\nu=11$ degrees of freedom.

The main result of the afterglow modeling is that the data favors a true jet energy of 
$E\sim(10^{48.5}-10^{49.5})\xi_{e,-1}^{-1}\;$erg and circumburst density $n_0\sim(10^{-5}-10^{-2})\xi_{e,-1}^{-1}$
for a viewing angle $\theta_{\rm obs}\sim0.28 - 0.45\;$rad ($16^\circ-26^\circ$). The forward shock micro-physical parameters 
admit values of $\epsilon_e\sim(10^{-1.7}-10^{-0.7})\xi_{e,-1}$ and $\epsilon_B\sim(10^{-5.6}-10^{-1.7})\xi_{e,-1}$ where the 
power-law index of the radiating electron distribution is $p\approx2.2$. 
We find that a non-thermal acceleration efficiency of $\xi_e\sim0.1$ corresponds to  reasonable inferred values for the 
afterglow model parameters. Broadly similar values of the model parameters were also obtained in recent works 
\citep{Hallinan+17,Kim+17,Margutti+17,Troja+17}.


\section {Discussion}
\label{sec:diss}

The new discovery of the first GW signal from the merger of a binary NS system, which was also accompanied by a SGRB, \grb, solidifies the role of binary NS mergers as the progenitors of SGRBs, and opens a new window in multi-messenger astronomy that enables us to learn more about SGRB physics. The great interest that this has raised in the community resulted in an exquisite follow-up campaign, which allows us to learn about the properties of the outflow generated in the course of this explosive event. The outflow appears to contain a sub- or mildly relativistic neutron-rich outflow that powers a kilonova-like emission accounting for the early IR to UV emission through radioactive decay,  as well as afterglow emission from a narrow relativistic jet that we are viewing off-axis from outside of its initial aperture.

In this work we have first briefly examined the IR, optical and UV emission from half a day up to about a week after the event (in \S\,\ref{sec:optical}). We concluded that its quasi-thermal spectrum as well as peak time and temperature favor a kilonova origin over other alternatives mentioned in the literature. Moreover, comparison of the data to kilonova models favors the formation of an HMNS (which is supported against collapse to a BH by differential rotation, and has an expected lifetime of $t_{\rm HMNS}\lesssim 1\;$s before collapsing to a BH), with a lifetime of $t_{\rm HMNS}\gtrsim300\;$ms. This is consistent with the fact that the GW signal that does not strictly rule out the formation of a short-lived HMNS after the 
NS-NS merger \citep{Abbott+17b}.
On the other hand, while the kilonova emission favors a long-lived ($\gtrsim300\;$ms) massive NS remnant, this remnant does not have to be a HMNS, and could instead possibly be a supra-massive NS (SNMS), which is supported against collapse to a BH by uniform rigid-body rotation, and therefore typically collapses to a BH only on the order of its spindown time due to magnetic dipole braking, which is typically $\gtrsim10^2\;$s even for a magnetar-strength magnetic field. A stable NS (even without rotational support) could in principle also be produced, if the remnant mass is not too large and if the equation of state (EOS) is rigid enough.

In \S$\,$\ref{sec:host} we used the measured offset of the optical/IR emission from the center of \host, of $r_\perp = 2\;$kpc, to constrain the distance $s = v_{\rm sys}t_{\rm mer}\lesssim2\,$--$\,10\;$kpc traveled by the NS-NS system from its birth to its merger location, assuming straight-line motion. Given the old stellar ages in elliptical galaxies, we find that one must account for the galaxy's gravitational potential and integrate the binary's possible trajectories in order to make more reliable quantitative inferences from $r_\perp$.

We have presented (in \S\,\ref{sec:model}) a simple analytic model for the peak time ($t_{\rm peak}$) and flux ($F_\nu^{\rm peak}$) of the afterglow emission from viewing angles outside of the jet's initial aperture ($\theta_{\rm obs}>\theta_0$), as well as the flux evolution after the peak, and provided provided the general the general predictions of this mode (for $\theta_{\rm obs}\gtrsim2\theta_0$; in \S\,\ref{sec:genres}). 

It was then contrasted with the results of numerical lightcurves from hydrodynamic simulations of the GRB jet during the afterglow phase (in \S\,\ref{sec:sims}). In the X-ray and optical, where the on-axis lightcurves decay at early times, the lightcurves from viewing angles mildly outside of the initial jet aperture, $\theta_0<\theta_{\rm obs}\lesssim(2-3)\theta_0$, were found to peak earlier than the analytic predictions (due to slower material on the sides of the jet and some lateral spreading, which cause the jet's beaming cone to reach such $\theta_{\rm obs}$ faster than the analytic expectations), while for larger $\theta_{\rm obs}$ the lightcurves peak later than the analytic expectation (since the jet decelerates and expands sideways more slowly with radius and observed time compared to the analytic models that feature an exponential lateral expansion with radius, and hence its beaming cone reaches large $\theta_{\rm obs}$ at later times). The analytic prediction for the peak flux is higher than that from numerical simulations by roughly an order of magnitude, 
since the numerical peak time occurs somewhat before the off-axis lightcurve join the on-axis lightcurve, while in the analytic model the two are assumed to coincide.

In the radio, since the on-axis lightcurve still rises when the beaming cone reaches the line of sight, the flux continues to gently rise, and it starts to decay only near the passage of $\nu_m$, (which is observed at somewhat different times for different viewing angles $\theta_{\rm obs}$), thus resulting in later peak times. The discrepancy in peak flux is smaller (see upper panel of Fig.~\ref{fig:LC_comp_off-axis}).

The numerical afterglow lightcurves from hydrodynamic simulations were directly contrasted with the data for \grb\ in
\S\,\ref{sec:compdata}. The data that was used included both upper limits and flux measurements, and the early IR to 
UV data were not included as they are likely dominated by the kilonova emission. In order to allow more efficient fits 
to the data given the reasonably large number of model parameters and corresponding parameter space to explore, we have 
used an approach that involves two different types of scalings \citep[following][]{scaling}: (i) scaling relations 
of the dynamical equations (an arbitrary rescaling of the energy and density) in order to avoid the need for performing 
a large number of hydrodynamic simulations, and (ii) additional scalings of the flux density at the different power-law 
segments of the afterglow synchrotron spectrum (with the shock microphysics parameters), in order to greatly reduce the 
number of required numerical lightcurve calculations.

We perform a non-linear least-squares fit of the afterglow model using the measured and upper-limit fluxes in X-ray and 
radio and find that the data can be explained by a modest $\theta_{\rm obs}\sim(1.4-2.25) \theta_0$, where the initial 
aperture of the jet $\theta_0=0.2=11.46^\circ$. Correspondingly, we find that the true jet energy 
$E\sim(10^{48.5}-10^{49.5})\xi_{e,-1}^{-1}\;$erg is on the higher end as compared to previous SGRB observations. Our model fits 
constrained the circumburst density to $n\sim(10^{-5}-10^{-1})\xi_{e,-1}^{-1}\;{\rm cm}^{-3}$ that agrees with the environment typically found 
in an elliptical galaxy. Even with such large total energy release, the relatively much smaller observed $E_{\gamma,\rm iso}\approx5.4\times10^{46}\;$erg provides excellent verification of the large off-axis angle favored by the afterglow data.

\vspace{0.5cm}
\section*{Acknowledgements}
We thank E. Sobacchi, E. Behar, and L. Yacobi for useful discussions.
JG and RG acknowledge support from the Israeli Science Foundation under Grant No. 719/14. 
RG is supported by an Open University of Israel Research Fellowship.


\end{document}